\documentclass[a4paper,fleqn]{mnras}

\usepackage{newtxtext,newtxmath}

\usepackage[T1]{fontenc}
\usepackage{ae,aecompl}

\usepackage{graphicx}	
\usepackage{amsmath}	

\newcommand{\lapp}{\ifmmode\stackrel{<}{_{\sim}}\else$\stackrel{<}{_{\sim}}$\fi}
\newcommand{\gapp}{\ifmmode\stackrel{>}{_{\sim}}\else$\stackrel{<}{_{\sim}}$\fi}

\newcommand{\fgl} {4FGL\,J2039.5$-$5617}
\newcommand{\msp} {PSR\,J2039$-$5617}

\title[Radio pulses from \msp]{Radio pulsations from the $\gamma$-ray millisecond pulsar \msp}

\author[A. Corongiu et al.]{
A. Corongiu$^{1}$\thanks{E-mail: alessandro.corongiu@inaf.it},
R. P. Mignani$^{2,3}$,
A. S. Seyffert$^{4}$,
C. J. Clark$^{5}$,
C. Venter$^{4}$,
L. Nieder$^{6,7}$,
\newauthor {A. Possenti$^{1,8}$,
M. Burgay$^{1}$,
A. Belfiore$^{2}$,
A. De Luca$^{2}$,
A. Ridolfi$^{1,9}$,
Z. Wadiasingh$^{10,11,4}$
}
\\
$^{1}$INAF - Osservatorio Astronomiico di Cagliari, Via della Scienza 5, I-09045 Selargius (CA), Italy\\
$^{2}$INAF - Istituto di Astrofisica Spaziale e Fisica Cosmica, Via Alberto Corti 12, I-20133 Milano, Italy\\
$^{3}$Janusz Gil Institute of Astronomy, University of Zielona G\'ora, ul Szafrana 2, 65-265, Zielona G\'ora, Poland \\
$^{4}$Centre for Space Research, North-West University, Potchefstroom Campus, Private Bag X6001, Potchefstroom, South Africa, 2520 \\
$^{5}$Jodrell Bank Centre for Astrophysics, Department of Physics and Astronomy, The University of Manchester, M13 9PL, UK\\
$^{6}$Albert Einstein Institut, Max Planck Institut f\"ur Gravitationsphysik, D-30167 Hannover, Germany\\
$^{7}$Leibniz Universit\"at Hannover, D-30167 Hannover, Germany\\
$^{8}$ Universit\'a degli Studi di Cagliari, Dipartimento di Fisica,
S.P. Monserrato$-$Sestu km 0,700, I-09042 Monserrato (CA), Italy\\
$^{9}$ Max Plank Institute f\"ur Radioastronomie, Auf dem H\"ugel 69, D-53121 Bonn, Germany \\
$^{10}$Astrophysics Science Division, NASA Goddard Space Flight Center, Greenbelt, MD 20771, USA\\
$^{11}$Universities Space Research Association (USRA) Columbia, MD 21046, USA
}

\date{Accepted XXX. Received YYY; in original form ZZZ}

\pubyear{2020}

\begin{document}
\label{firstpage}
\pagerange{\pageref{firstpage}--\pageref{lastpage}}
\maketitle
\begin{abstract}
The predicted nature of the candidate redback pulsar
3FGL\,J2039.6$-$5618 was
recently confirmed by the discovery of $\gamma$-ray millisecond
pulsations (Clark et al. 2020, hereafter
 Paper\,I), which identify this $\gamma$-ray source as \msp. We
observed 
this object with the Parkes radio telescope in 2016 and 2019. We
detect radio pulsations at 1.4\,GHz and 3.1\,GHz, at the 2.6ms
period discovered in $\gamma$-rays, and also at 0.7\,GHz in
one 2015 archival observation. In all bands, the radio pulse profile
is characterised by a single relatively broad peak which leads the
main $\gamma$-ray peak. At 1.4\,GHz we found clear evidence of
eclipses of the radio signal for about half of the orbit, a
characteristic phenomenon in redback systems, which we associate with
the presence of intra-binary gas.  From the dispersion measure
of $24.57\pm0.03$\,pc\,cm$^{-3}$ we derive a pulsar distance of
$0.9\pm 0.2$\,kpc or $1.7\pm0.7$\,kpc, depending on the assumed
Galactic electron density model.
The modelling of the radio and $\gamma$-ray light curves
leads to an independent determination of the orbital
inclination, and to a determination of the pulsar mass, qualitatively
consistent to the results in  Paper\,I.
\end{abstract}




\section{Introduction}

Millisecond pulsars (MSPs) differ from the bulk of the
rotation-powered pulsar population in that their spin periods are much
shorter ($P_{\rm s}\la$10\,ms) and their spin-down rates much smaller
($\dot{P_{\rm s}} \sim 10^{-21}-10^{-18}$) s s$^{-1}$. This implies
that they have large characteristic ages ($P_{\rm s}/2\dot{P_{\rm
s}}\sim$1--10 Gyr), making them the oldest pulsars, with the
lowest dipolar magnetic fields ($B \sim 10^{8}-10^{9}$ G), although
with still high rotational energy loss rates ($\dot{E}\sim
10^{32}-10^{36}$\,erg\,s$^{-1}$). The short spin periods suggest
that they have been spun up by accretion from a companion star,
according to the standard ``recycling scenario'' (Alpar et al.\ 1982;
Radhakrishnan \& Srinivasan 1982). Accordingly, about two thirds ($\sim
210$\footnote{ATNF pulsar catalogue v1.60 (Manchester et al.\ 2005)}) of
known
MSPs are in binary systems , usually with white dwarf (WD)
companions, either a He WD of mass $0.1 M_{\odot} \la M_{\rm C} \la
0.5 M_{\odot}$ or a Carbon-Oxygen (CO) WD, of mass $0.5 M_{\odot} \la
M_{\rm C} \la 1 M_{\odot}$, with orbital periods of up to hundreds of
days (Hui et al.\ 2018). Some binary MSPs, however, have
non-degenerate low-mass companions and very short orbital periods
($P_{\rm orb}<$1 d). These are known as ``Redbacks'' (RBs; Roberts
2011), with companion mass $M_{\rm C}\sim$0.1--0.4$M_{\odot}$ which
are partially ablated by irradiation from the pulsar wind. RBs are
related to the classical ``Black Widows'' (BWs; Fruchter et
al.\ 1988), binary MSPs which have lighter companions of mass $M_{\rm
  C}\la0.05M_{\odot}$, that are almost fully ablated. These ``spiders''
are ideal systems with which to study the MSP recycling process, the
acceleration, composition and shock dynamics of the MSP winds, and the
possible formation of isolated MSPs via full ablation of the companion.
They are also  excellent targets for MSP mass measurements (e.g., van
Kerkwijk et al.\ 2011), key to determining the neutron star equation of
state. Whether RBs and BWs are
linked by evolution or they represent two different channels of the
binary MSP evolution is still debated (e.g., Chen et al.\ 2013).

Apart from the radio and optical bands, where only the companion
star is usually detected, binary MSPs are also observed at high
energies (see, Torres \& Li 2020 for a review). In $\gamma$ rays, the
{\em Fermi} Large Area Telescope (LAT; Atwood et al.\ 2009) has detected
about 90 binary MSPs in the Galactic
field\footnote{{\tt https://confluence.slac.stanford.edu/display/GLAMCOG/\\Public+List+of+LAT-Detected+Gamma-Ray+Pulsars}},
about twice as many as those detected in the X-rays (Lee et
al.\ 2018). Whereas their $\gamma$-ray emission is mostly ascribed to
emission processes from within the pulsar magnetosphere, the X-ray
emission can originate either from the MSP itself (magnetosphere or
heated polar caps) or from the intra-binary shock formed by the
interaction of its wind and gas from the ablated companion, like
in RBs and BWs (Harding \& Gaisser 1990, Arons \& Tavani 1993, Roberts
et al.\ 2014). Observations in the $\gamma$-ray energy band have proved to
be instrumental to the discovery of new binary MSPs, especially of new
BWs/RBs which are elusive targets in radio pulsar surveys owing to
partial eclipses of the radio beam caused by the intra-binary plasma
from the companion star ablation, a process that does not
affect the propagation of the $\gamma$-ray emission beam. Indeed, only
a handful of such ``spiders'' were known before the advent of {\em
  Fermi} and the quest for new BWs/RBs among unidentified {\em
  Fermi}-LAT sources is restlessly pursued (e.g., Hui 2014; Hui \& Li
2019), with promising candidates for radio/$\gamma$-ray pulsation
searches selected by machine-learning algorithms (e.g., Saz Parkinson
et al.\ 2016).  In many cases, optical observations have been
instrumental to such searches through the discovery of the orbital
period from the detection of $<1$ d periodic modulations in the flux
of the putative companion star like, e.g. for the BW PSR\,
J1311$-$3430 (Romani 2012; Pletsch et al.\ 2012; Ray et al.\ 2013) and
the RB PSR\, J2339$-$0533 (Romani \& Shaw 2011; Ray et al.\ 2014).

As of now, 43 BW/RB candidates have been confirmed as radio/$\gamma$-ray
pulsars in the Galactic field, while 11 still lack detected pulsations
(Linares 2019). One of the latter is the {\em Fermi} source 3FGL\,
J2039.6$-$5618, now \fgl\footnote{Hereafter in this work we use the
name \fgl\ when referring to this object as a {\em Fermi} source, and the
name \msp\ when referring to it as a pulsar.} (Fermi Large Area Telescope
Fourth Source Catalog, 4FGL; Abdollahi et al.\ 2020), singled out as an
RB candidate based upon the detection of
a periodic flux modulation ($\sim$0.22 d) in its X-ray/optical
counterpart with {\em XMM-Newton} and GROND at the MPG 2.2m telescope
(Salvetti et al.\ 2015; Romani 2015).  In addition, possible evidence
of a $\gamma$-ray modulation has been found in the {\em Fermi} data
(Ng et al.\ 2018). 

Recently, the radial velocity curve of the \fgl\ 
counterpart has been measured through optical spectroscopy (Strader et
al.\ 2019), confirming that the period of the optical flux modulations
discovered by Salvetti et al.\ (2015) and Romani (2015) indeed
coincides with the orbital period of a tight binary system.  The
improved measurement of the orbital period and the determination of
the binary system's orbital parameters were used to perform a targeted
search for $\gamma$-ray pulsations in the {\em Fermi}-LAT data. The
resulting detection of $\gamma$-ray pulsations at a period of 2.6 ms (Clark et al.\ 2020,
hereafter  Paper\,I) confirmed the MSP identification of \fgl\ 
making it the third BW/RB directly identified in
$\gamma$-rays after PSR\, J1311$-$3430 (Pletsch et al.\ 2012).
and PSR\,J1653-0158 (Nieder et al.\ 2020).
\fgl\ 
(now PSR\, J2039$-$5617) has not yet been detected as
an X-ray pulsar owing to the lack of suitable observations with either
{\em XMM-Newton} or {\em Chandra}. It also eluded detection in
previous radio observations (Petrov et al.\ 2013; Camilo et
al.\ 2015).  Here, we report on the first detection of radio
pulsations from \fgl\ 
using more recent observations
that we obtained with the Parkes radio telescope in 2016, before the
source had been identified as a $\gamma$-ray MSP.

This  paper is structured as follows. In Section\,\ref{sec:obs} we
describe the radio observations. The data analysis and results
are presented in Section\,\ref{sec:analysis} and discussed in
Section\,\ref{sec:discussion}, respectively. The summary follows in
Section\,\ref{sec:summary}.

\section{Observations}
\label{sec:obs}

We observed \fgl\ between May and September 2016, prior to the detection of $\gamma$-ray pulsations, with the Parkes radio
telescope to search for radio pulsations and confirm its
proposed identification as a binary MSP (Salvetti et
al.\ 2015). We
pointed the telescope at the most recent $\gamma$-ray coordinates of
\fgl\ at the time of the observations according to the Fermi Large
Area Telescope Third Source Catalog (3FGL; Acero et al.\ 2015),
RA$_{\rm J2000}=20^{\rm h} 39^{\rm m} 40\fs32$, DEC$_{\rm
  J2000}=-56^\circ 18\arcmin 43\farcs6$\footnote{The updated 4FGL
  coordinates, RA$_{\rm J2000}=20^{\rm h} 39^{\rm m} 35\fs4$,
  DEC$_{\rm J2000}=-56^\circ 17\arcmin 01\farcs0$, fall within the
  observed field of view. }. We observed the source with the central
beam of the Multi-beam receiver (central frequency $\nu_{\rm
  c}=1390$\,MHz, band-width BW$=256$\,MHz), and with the high
frequency feed of the coaxial 10$-$40 dual band receiver ($\nu_{\rm
  c}=3100$\,MHz, BW$=1024$\,MHz). The source signal was
digitised and recorded in pulsar search mode by the PDFB4 backend. 
Neither flux nor polarization calibration have been done since such
calibration has no impact on the pulse shape.
The number of frequency channels and bits per sample are reported in
Table\,\ref{tab:obslog}, where we present a summary of all
observations, both ours and archival (see below in this section),
discussed in this work. Technical details about the instrument can be
found on the Parkes telescope web page\footnote{{\tt
    https://www.parkes.atnf.csiro.au}}, and references therein.

\begin{table*}
  \centering
  \caption{\fgl\,Observation log. Symbols are defined as follows.
    T$_{\rm obs}$: observation length; $\nu_{\rm c}$: central
    frequency of the acquired frequency band; $\Delta\nu$: frequency
    band witdh; N$_{\rm bit}$: number of bits per sample; N$_{\rm
      chan}$: number of frequency channels; t$_{\rm sampl}$: sampling
    time; Orbital coverage: orbit's phase range covered by the
    observations.}
  \label{tab:obslog}
  \begin{tabular}{lrrrcrrc}
    \hline
    Date & T$_{\rm obs}$ & $\nu_{\rm c}$ & $\Delta\nu$ & N$_{\rm bit}$ &
    N$_{\rm chan}$ & t$_{\rm sampl}$ & Orbital coverage \\
    & seconds & MHz & MHz & & & $\mu$s & phase range \\
  \hline
    2015 Apr 9  &  3605.8 &  732 &   64 & 2 &  512 &  64 & 0.12$-$0.30\\
    2015 Apr 9  &  3605.8 & 3100 & 1024 & 2 &  512 &  64 & 0.12$-$0.30\\
    2015 Apr 12 &  3605.8 &  732 &   64 & 2 &  512 &  64 & 0.43$-$0.61\\
    2015 Apr 12 &  3605.8 & 3100 & 1024 & 2 &  512 &  64 & 0.43$-$0.61\\
    2016 May 8  & 21133.1 & 3100 & 1024 & 1 &  512 & 144 & ALL\\
    2016 May 24 & 21133.1 & 1369 &  256 & 1 &  512 & 144 & ALL\\
    2016 July 6  &  7205.7 & 3100 & 1024 & 2 &  512 & 200 & 0.76$-$0.12\\
    2016 Aug 19 &  5412.9 & 1369 &  256 & 2 & 1024 & 200 & 0.29$-$0.57\\
    2016 Sept 9  & 11776.8 & 1369 &  256 & 2 & 1024 & 200 & 0.24$-$0.84\\
    2019 June 19 & 15933.6 & 1369 &  256 & 4 & 1024 & 124 & 0.46$-$0.27 \\
    2019 June 20 & 14417.9 & 1369 &  256 & 4 & 1024 & 124 & 0.74$-$0.48\\
    \hline
  \end{tabular}
\end{table*}

The choice of observing \fgl\ at two different frequencies was
grounded on the difficulties of observing pulsars in RB systems.
The presence of intra-binary gas in RB systems leads to orbital phase
dependent and variable signal absorption and dispersion, phenomena
that make the detection of pulses more difficult. Observations at higher
frequencies, where these effects are less severe,
can therefore be beneficial. On the other hand, given the typical
radio pulsar power-law (PL) spectrum $S(\nu)\propto\nu^{\alpha_{\nu}}$,
where the distribution for the spectral index $\alpha_{\nu}$ ranges from -3.5 to +1.5 and peaks at -1.57 (e.g.,
Jankowski et al.\ 2018), pulsars
appear brighter at lower frequencies. Observing at lower frequencies
therefore allows pulses to be detected with a higher flux when the pulsar
is at orbital phases around inferior conjunction, i.e. in front of the
intra-binary gas cloud when seen from the observer.  Therefore,
we chose to observe \fgl\ in two bands to improve the detection
chances.
 
At both frequencies, our strategy consisted of observing \fgl\ for one
entire orbit ($\sim 5.3$\,hr), and twice for about one quarter of
an orbit around inferior conjunction. We computed the orbital phases
on the basis of the orbital period $P_{\rm B}=0.22748\pm0.00043$\,d
and epoch of the ascending node $T_{\rm asc}=56884.9667\pm0.0003
\,{\rm MJD}$ determined from observations of the optical flux
modulations by Salvetti et al.\ (2015), which were the most accurate
reference values at the time our proposal was submitted. One of the
two planned short observations at 3.1\,GHz could not be executed
because a technical problem occurred at the telescope and was not rescheduled.  Therefore, only five of the six planned observations
were executed. Under Director Discretionary Time, we carried out
complementary follow-up radio observations of \fgl\ from Parkes on
2019 June 19 and 20, which cover nearly one
entire orbit each.  These new observations were motivated by the
detection of radio pulsations from the analysis of our 2016 data (see
Sec. \ref{subsec:detection}) using a preliminary $\gamma-$ray timing
ephemeris described in Paper\,I, and were obtained in preparation of a
regular monitoring campaign of this source from Parkes.
The submitted proposal, ATNF project code
P1025 (PI. Corongiu), has already been accepted at the time of writing
and the related observations have been scheduled during the October
2019--March 2020 semester. This time, we used the Ultra-Wideband Low
(UWL, Hobbs et al.\ 2019) receiver that allows one to observe in the
frequency band $0.7$--$4.0$\,GHz. The source signal has been
digitised and recorded in pulsar search mode by the PDFB4 backend
for the 256\,MHz band centered at 1.4\,GHz.
  
Finally, in this work we also revisited public Parkes radio data,
available at the Commonwealth Scientific and Industrial Research
Organisation (CSIRO) Data Access Portal\footnote{\tt
  https://data.csiro.au/dap} (DAP), taken in two observing sessions on
2015 April 9 and 12 with both bands of the
10--40 receiver, and acquired with the PDFB3 (0.7\,GHz, $\nu_{\rm
  c}=732$\,MHz, BW$=64$\,MHz) and PDFB4 (3.1\,GHz) backends,
respectively. These are the data taken by Camilo et al.\ (2015) in
their radio survey of unidentified \textit{Fermi}-LAT sources, which we analysed to search a
posteriori for the radio pulsations detected in our 2016 observations
using the $\gamma$-ray timing ephemeris ( Paper\,I).

Therefore, this work presents a complete summary of the radio
observations of \fgl\ to date, all performed with the Parkes telescope, and spanning nine different epochs from 2015 to 2019.
 
\section{Data Analysis and Results}
\label{sec:analysis}

\subsection{Pulse search and detection}
\label{subsec:detection}

Search mode data for all observations have been phase-folded using the
routine {\tt dspsr}\footnote{{\tt http://dspsr.sourceforge.net}}. Folded
archives were created for each observation separately, with
sub-integrations of ten seconds and 64 pulse phase bins, maintaining the
same number of frequency channels as the raw data
(Table\,\ref{tab:obslog}). As seen in  Paper\,I, the orbital period of the
system varies significantly over time. We therefore used the $\gamma$-ray
ephemeris from Paper\,I to interpolate the orbital period and find the
epoch of the closest ascending node passage to each observation.

The presence of radio frequency interferences (RFI) is a known problem in
the radio data, therefore
we visually inspected each folded archive with the routine {\tt pazi},
provided by the software suite {\tt psrchive\footnote{{\tt
    http://psrchive.sourceforge.net}}}, which
allows one to graphically select and remove unwanted
channels/sub-integrations and to check the resulting
integrated profile at runtime.

Since the ephemeris reported in Paper\,I was obtained from
$\gamma$-ray observations, the radio dispersion measure (DM) remained
unknown. Without de-dispersion, the pulse remained undetected in a
preliminary visual inspection of the Parkes data. We therefore ran a DM
search on each archive separately, by processing them with the routine
{\tt pdmp} provided by the suite {\tt psrchive}. This routine searches for the
best spin period and DM in a pre-defined grid of values, using a
selection criterion based on the highest signal--to--noise (S/N) ratio
of the integrated profile, i.e. the profile obtained by summing in
phase the pulse profiles of all sub-integrations and all frequency
channels.

We performed the DM search in two steps. The first step consisted of a
coarse search along the entire range of DM values predicted for a pulsar
lying within the Galaxy, and along the line-of-sight (LOS) to \fgl. The
maximum DM value (DM$_{\rm max}$) was obtained from the two most recent
models for the free electron distribution in the Galaxy.  At the Galactic
coordinates of \fgl,
$l=341\fdg230863$, $b=-37\fdg154895$, the NE2001 (Cordes \& Lazio 2002)
model predicts DM$_{\rm max}=53.55$\,pc\,cm$^{-3}$, while the YMW16 (Yao
et al.\ 2017) model predicts DM$_{\rm max}=38.81$\,pc\,cm$^{-3}$. We
adopted a very conservative approach with respect to these predictions
and sampled DM values up to 80\,pc\,cm$^{-3}$, with a DM step of
0.01\,pc\,cm$^{-3}$.

This first coarse search already allowed us to identify the
observations where radio pulsations are detected, and to obtain an initial estimate for the DM. After de-dispersing the archives
with the initial DM estimate, we ran a second more refined
search with a half range of 0.25\,pc\,cm$^{-3}$ and step of
0.001\,pc\,cm$^{-3}$. Pulsations were visually recognised in the same
observations identified in the first run of our DM search.  The best
DM value for each observation is reported in the last column of
Table\,\ref{tab:detections}, where we present a summary of the outcome
of our search for pulses. The average value is
DM$=24.57\pm0.03$\,pc\,cm$^{-3}$, which we will use throughout this
work.

\begin{table*}
  \centering
  \caption{\fgl\, radio pulses detection summary. See
    \S\,\ref{subsec:eclipses} for a discussion about the pulse
    detection along the whole orbit. Symbols are defined as
    follows. $\nu_{\rm c}$: central
    frequency of the acquired frequency band; S/N: signal--to--noise ratio; T$_{\rm det}$: time span of
    pulses effective detection; BW$_{\rm det}$: width of the frequency
    band along which pulses are effectively detected; Orbital detection: orbital phase range where
    pulses have been detected - values marked with a '*'
    indicate observed beginnings and ends of signal eclipses; Flux:
    derived flux density;  DM: dispersion measure.}
  \label{tab:detections}
  \begin{tabular}{lcccccccc}
    \hline
    Date & $\nu_{\rm c}$ & Detected & S/N & T$_{\rm det}$ &
    BW$_{\rm det}$ & Orbital detection &
    Flux & DM \\
    & MHz & & & seconds & MHz & phase range & mJy & pc\,cm$^{-2}$  \\
    \hline
    2015 Apr 9 & 732 &  NO & $-$   &  $-$  &  $-$ & NONE                    & $-$  & $-$              \\
    2015 Apr 9 & 3100 & NO & $-$   &  $-$  &  $-$ & NONE                    & $-$  & $-$              \\
    2015 Apr 12& 732 & YES & 17.48 &  1440 &   10 & 0.55*$-$0.61       & 1.94 & 24.453$\pm$0.047 \\
    2015 Apr 12& 3100 & NO & $-$   &  $-$  &  $-$ & NONE                    & $-$  & $-$              \\
    2016 May 8 & 3100 & YES & 26.15 & 21133 & 1024 & ALL
    & 0.07 & 24.384$\pm$0.230 \\
    2016 May 24& 1369 &YES & 52.18 &  9360 &   80 & 0.55*$-$0.05* & 0.56 & 24.745$\pm$0.075 \\
    2016 July 6 &  3100 &NO & $-$   & $-$   &  $-$ & NONE                    & $-$  & $-$              \\
    2016 Aug 19&  1369 & NO & $-$   & $-$   &  $-$ & NONE                    & $-$  & $-$              \\
    2016 Sept 9 & 1369 &YES & 19.25 &  4800 &   96 & 0.58*$-$0.84       & 0.27 & 24.742$\pm$0.081 \\
    2019 June 19& 1369 & YES & 88.01 &  9840 &  198 & 0.50*$-$1.00* & 0.59 & 24.676$\pm$0.080 \\
    2019 June 20& 1369 & YES & 57.53 &  3600 &   94 & 0.74$-$0.93*       & 0.92 & 24.482$\pm$0.080 \\
    \hline
  \end{tabular}
\end{table*}

After implementing the DM value in each archive where pulses have been
detected, the pulse profiles seen in separate sub-integrations did not
perfectly align, the only exception being the 3.1\,GHz observation on
2016 May 5. The misalignment behaviour was consistent with a
linear trend with respect to the pulse phase. These small misalignments
are most likely caused by the uncertainty in the orbital phase predicted
by the $\gamma$-ray ephemeris, as this can be measured on single epochs 
with higher precision in the radio data. There may also be small
phase deviations due to the assumed astrometric parameters, which were
fixed at the {\em Gaia} DR2 solution ({\em Gaia} Collaboration 2016,
2018) without further refinement. Therefore, for each individual
observation, we extracted a set of pulse's times of arrival (ToAs), whose
number depended on the
pulsar brightness and the duration of the time interval where pulses
were visible (see, Table\ref{tab:detections}). Using the obtained ToAs,
we used each observation's optimum DM value as determined above, and fit
for the epoch of the ascending node and the pulsar spin period, the two
parameters that the sub-integration alignment is most sensitive
to, using the $\gamma$-ray ephemeris as a starting solution. The obtained
values for the spin period and the epoch of the
ascending node were then implemented in the archives to obtain
the best possible alignment of the pulse profiles along the
sub-integrations. In the 2015 April 9 0.7\,GHz and 2016
September 9 1.4\,GHz observations the pulsar flux was too
low and pulses were visible for too a short time to obtain reliable
results from this procedure. For these two observations we obtained
the best possible alignment by using the
value for the spin period given by the routine {\tt pdmp}, without
correcting the epoch of the ascending node.

Pulsations from \fgl\ were detected in six out of the eleven
observations, namely in the 2015 April 9 observation at
0.7\,GHz, in the 2016 May 8 observation at 3.1\,GHz, and in
four observations at 1.4\,GHz on 2016 May 24, September
9, 2019 June 19 and 2019 June 20.  We comment on the
lack of detection in the remaining observations in
\S\,\ref{subsec:eclipses} and \S\,\ref{subsec:scint}. The independent
detection of pulsations in six observations and
in three different frequency bands clearly demonstrates that \fgl\ is
also a radio pulsar and from now on we refer to it as \msp, adopting
the $\gamma$-ray pulsar name from  Paper\,I.

After the detection of a new radio pulsar, a timing analysis is usually
performed on the available data. Generally speaking, pulsar timing in
radio can achieve a significantly higher precision than what is
obtainable
with $\gamma$-ray data, but only if the available radio data has a high
cadence and covers a comparable epoch interval to the {\em Fermi}-LAT data.
The available radio data on \msp\ are too few and too sparse
to measure the global timing parameters with a precision similar to the
one achieved by
the $\gamma$-ray timing (Paper\,I), and they would have completely missed
the non monotonic variations of the orbital period (Paper\,I). Moreover,
radio pulses are not detectable at all orbital phases (see
\S\ref{subsec:eclipses}), and this affects the precision in the
measurement of the orbital parameters. For these reasons we did not
perform a timing analysis of the radio data alone and in particular for
the latter one we do not have plans for a timing campaign in the radio
domain.

\subsection{Pulse Profile Analysis}
\label{subsec:profs}

\begin{figure*}
\centering
\includegraphics[angle=0,width=\textwidth]{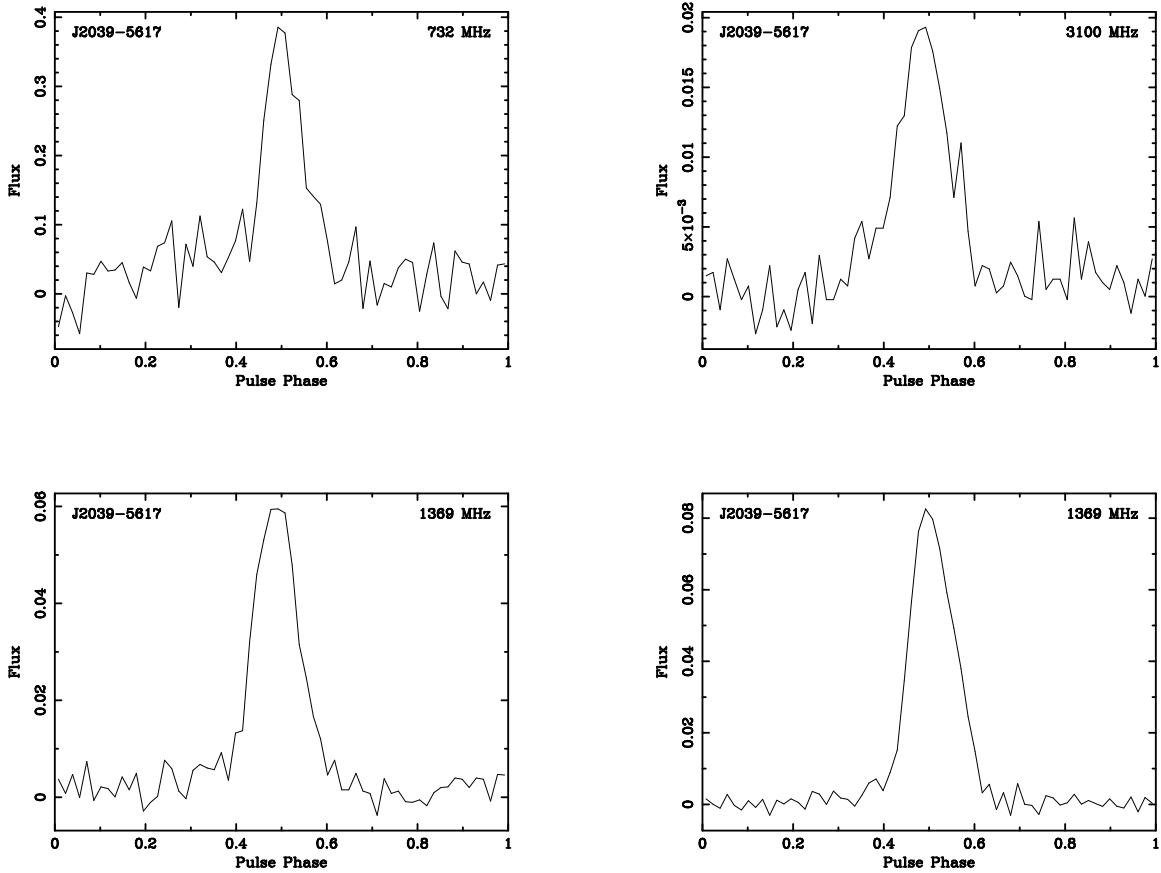}
\caption{Integrated pulse profiles of \msp\ at 0.7 (upper
  left panel), 3.1 (upper right panel) and 1.4\,GHz (lower panels).
  The 0.7\,GHz profile has been obtained from the 2015 April 12 observation after removing from the data sub-integrations and
  channels where the pulse is not detected. The 3.1\,GHz profile
  corresponds to the 2016 May 8 observation, the only one
  where pulses have been detected in this band. The 1.4\,GHz profiles
  in the lower panels have been obtained by summing the 2016 (May
  24, September 9, left panel) and the 2019
  (June 19, 20 right panel) observations,
  respectively. In all panels the peak has been displaced to phase 0.5
  for clarity.
  \label{fig:stds}}
\end{figure*}

Figure\,\ref{fig:stds} displays the integrated pulse profiles obtained
from the data of the 2015 0.7\,GHz observation, the single 2016
3.1\,GHz observation, the two 2016 1.4\,GHz observations, and the two 2019
1.4\,GHz observations. The displayed 1.4\,GHz pulse profiles have been
obtained by coherently adding in phase the profiles of the single
observations of the same year.  In all panels, we intentionally
displaced the profile to phase 0.5 for better clarity. In all cases,
the profile shape is typical of a single-peak pulse, consistent with a
polar cap emission, and the equivalent width at all frequencies is
$\sim0.1$ in pulse phase. The 0.7\,GHz and 3.1\,GHz profiles show some
additional features that are most likely due to the low S/N. Whether
these features are real or simply due to noise could be clarified with
additional observations in the future, coherently added in phase to increase the profile S/N.

\subsection{Pulse Brightness Analysis}

\subsubsection{Pulse brightness vs time: Signal eclipses}
\label{subsec:eclipses}

\begin{figure*}
  \centering
  \includegraphics[angle=0,width=\textwidth]{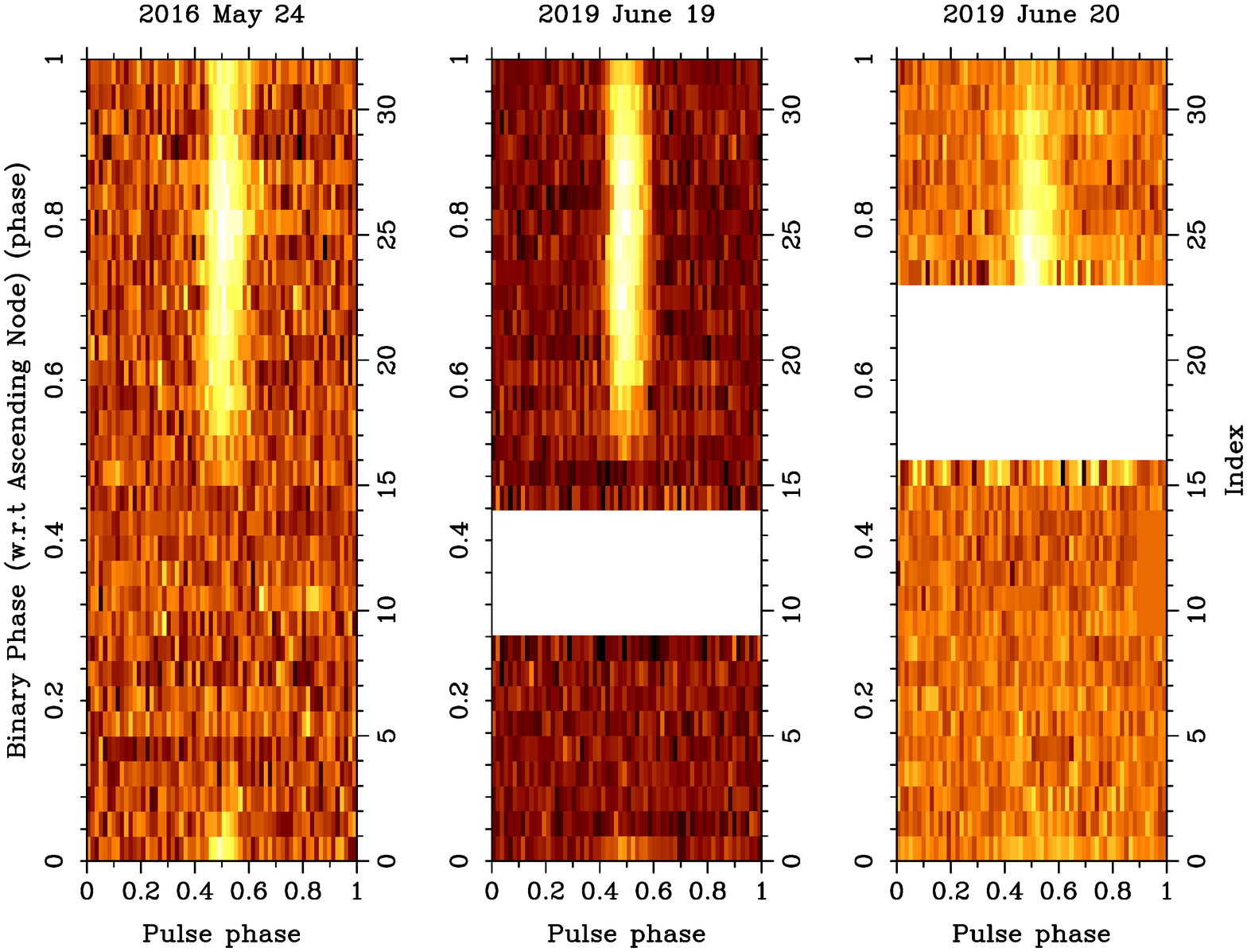}
  \caption{Colour map of the pulse amplitude vs.\ spin and orbital
    phase (horizontal and vertical axis, respectively) for the 1.4\,GHz
    2016 May 24 (left panel), 2019 June 19 (mid panel) and 2019
    June 20 (right panel) observations, which have the largest
    orbital phase coverage. In the mid and right panels the horizontal
    white band marks the orbital phase ranges that have not been
    covered by the observations. In all panels the scale on the left-hand
    side vertical axis is the
    orbital phase and that on the right one is the sub-integration number
    (Index). The occurrence of a signal eclipse is apparent in all
    three panels, although the orbital phase range (0.1--0.5) defined
    by the eclipse start and end is uniformly covered only by the
    first observation.   \label{fig:fullorbitsL}}
\end{figure*}

\begin{figure*}
  \centering
  \includegraphics[angle=0,width=\textwidth]{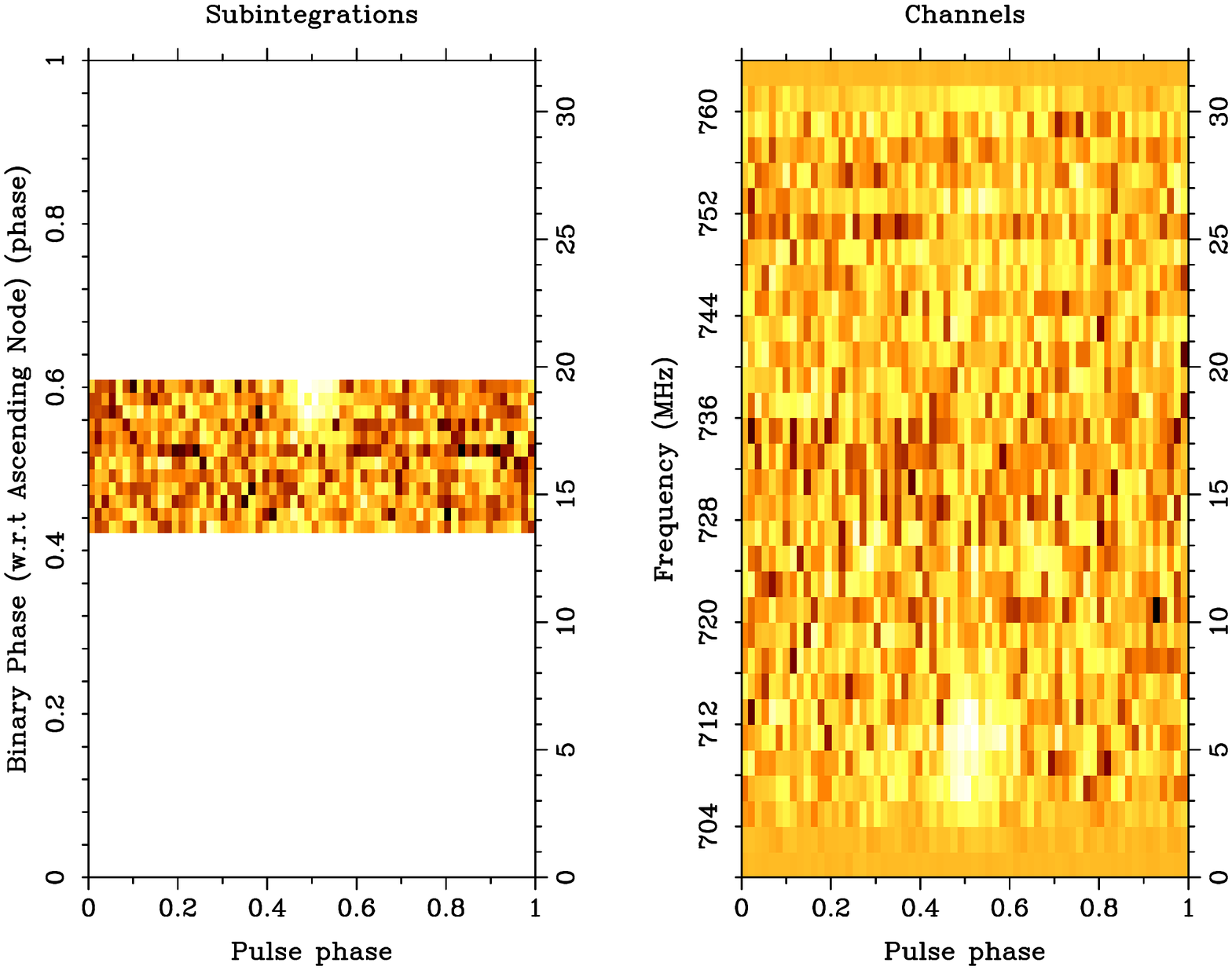}
  \caption{Colour map of the pulse amplitude vs.\ spin and orbital
    phase (left panel), and frequency (right panel) for the 2015 April
    12 observation at 0.7\,GHz.
    \label{fig:Pband-plots}}
\end{figure*}

The range of orbital phases spanned by each observation (see column 6
in Table\,\ref{tab:detections}) has been computed from the interpolated $\gamma-$ray ephemeris for each observation (see
\S\,\ref{subsec:detection}), after fitting for the epoch of ascending node in each, as described above. The
observations at 1.4\,GHz where pulses are detected
(Figure\,\ref{fig:fullorbitsL}) show that the pulsar signal is
eclipsed in the half orbit around superior conjunction ($\phi_{\rm
  orb}\sim0.25$). The orbital phase range where pulses are detected at
0.7\,GHz (Figure\,\ref{fig:Pband-plots}, left panel) is consistent
with this picture.  In these observations the edges of the signal's
eclipses have also been observed, and their orbital phases are
marked with an '*' in column 6 of Table \,\ref{tab:detections}. The
orbital phases of the beginning and end of the signal eclipses are not
stable from one orbit to another, as is commonly observed in other
RB systems, e.g. PSR\,J1740-5340A in the globular cluster
  NGC\,6397 (D'Amico et al.\ 2001) or PSR\,J1701-3006B in NGC\,6266
  (Possenti et al.\ 2003), and span an orbital phase range of
$\Delta\phi_{\rm orb}\sim0.1$.  The non stability of the orbital
phases at which signal eclipses begin and end at 1.4\,GHz is illustrated in Figure\,\ref{fig:fullorbitsL}. The colour map shows the signal amplitude as a function of pulse and orbital
phases for the three observations at this frequency that cover a
significant fraction of the orbit, namely the 2016 May 24
(left panel, 100\% of the orbit), the 2019 June 19 (mid
panel, 81\%) and June 20 (right panel, 74\%) observations.

The edges of the eclipse do not show any evidence of pulse delay or
broadening. Such a relatively sharp disappearance of the signal can be
ascribed to either a true occultation by the companion, or the
presence of intra-binary gas, either very hot or very cold. Indeed,
the orbital phase extent of the eclipse is undoubtedly at odds with
the occultation scenario, since this would require that the pulsar is
in a nearly surface-grazing orbit around the star, hence with an
orbital period much shorter than observed. Moreover, a search for
signal eclipses in the $\gamma-$ray data (Paper\,I) ruled out eclipses lasting longer than 0.1\% of an orbital period (about 20
seconds), suggesting that the star does not ever properly occult the
pulsar. The intra-binary gas scenario is instead an explanation that
also confirms the RB classification for \msp,
initially proposed on
the basis of optical observations (Salvetti et al.\ 2015; Strader et
al.\ 2019).

\begin{figure*}
\centering \includegraphics[angle=0,width=\textwidth]{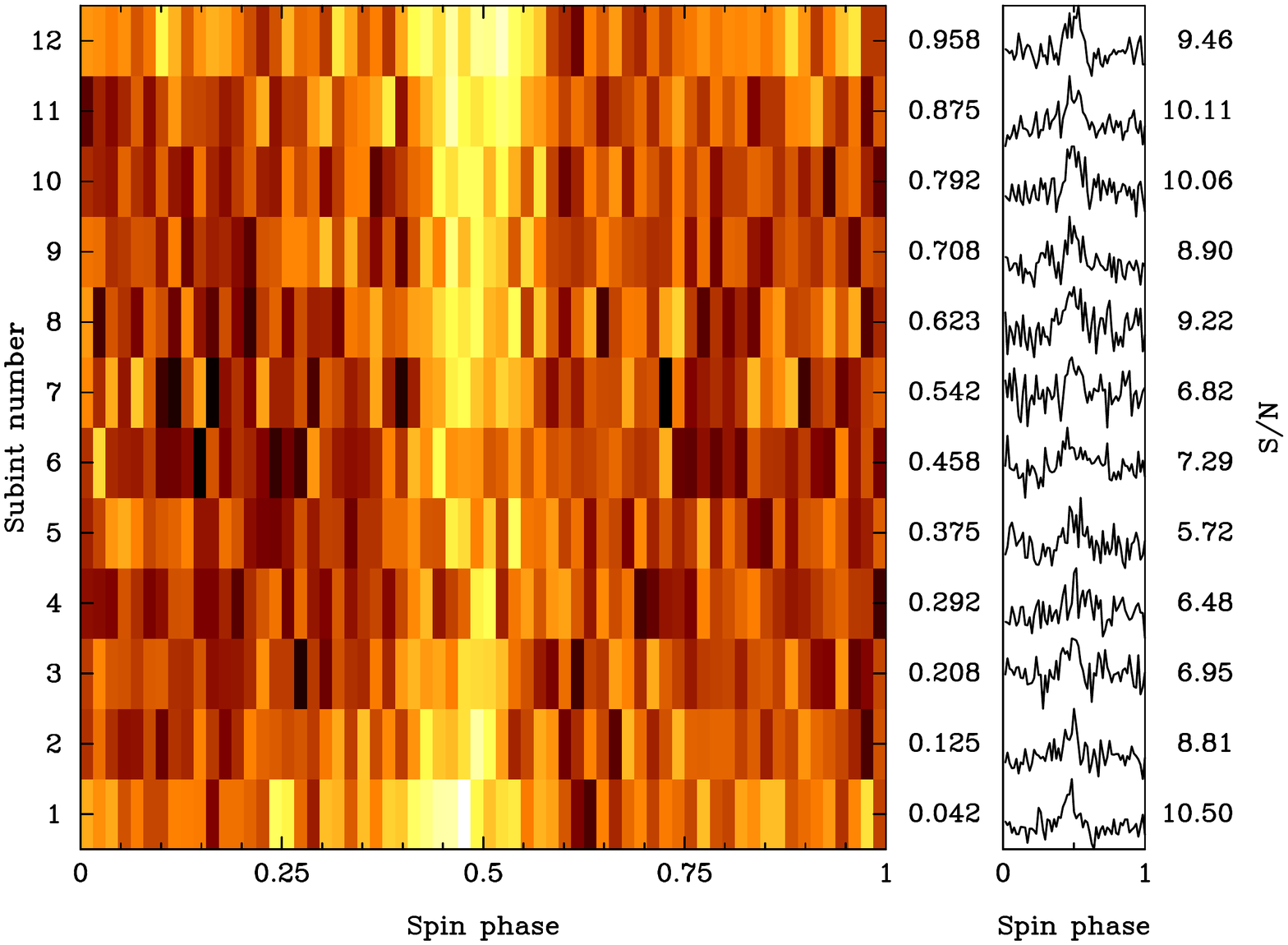}
\caption{Left panel: colour map of the pulse amplitude vs.\ spin and
  orbital phase (horizontal and vertical axis, respectively) for the
  2016 May 8 3.1\,GHz full-orbit observation. Right panel:
  stack plot of the pulse profile of each sub-integration in the left
  panel, whose S/N are reported on the right. The scale between the
  two panels indicates, for both panels, the mean orbital phase of
  each sub-integration/profile. The relatively high S/N level across
  all profiles seems to suggest that the radio signal is not eclipsed
  across the entire 3.1\,GHz (see discussion in \S
  \ref{subsec:eclipses}).
\label{fig:SBorb}}
\end{figure*}

The occurrence of signal eclipses at 3.1\,GHz can neither be confirmed
nor ruled out. Figure\,\ref{fig:SBorb} displays the signal amplitude as a
function of pulse and orbital phases for the 2016 May 8 full orbit
observation. The right-hand panel shows the pulse profiles of each
sub-integration from the left-hand panel. For each profile in
the right panel the mean orbital phase (left-hand side scale) and
S/N (right-hand side scale) is also reported. It is not clear, from visual inspection of the left
panel, whether or not the pulse remains detectable around superior conjunction. Moreover, no single pulse profile appears to feature the
typical {\em pure noise} profile, and for those profiles where the
pulsations seem less evident the corresponding S/N is not low enough
to firmly rule out the detection of a pulse. If eclipses do indeed also occur at 3.1\,Ghz, the current data suggests that they would occur
around the same orbital phase as at 1.4\,GHz but their duration would
be much shorter. This behaviour would be in line with what is observed
in other eclipsing radio pulsars, e.g. PSR\,J1748-2446A in the globular
cluster Terzan 5 (Rasio et al.\ 1991). Therefore, the
physical origin of possible signal eclipses at 3.1\,GHz is most likely
the same as at 1.4\,GHz: the non-detection of pulses is due to a signal
absorption which is less effective as the frequency increases, since the
optical depth inversely scales with the frequency, in some cases
$\propto\nu^{-0.4}$ (see, e.g., Broderick et al.\ 2016; Polzin et al.\
2018).

The phenomenology described above also allows us to shed some light on
the observations where pulses have not been detected. The archival
observation on 2015 April 9 and our own on 2016 August 19
were carried out when the pulsar was in an orbital phase
range where the signal is not detected in other observations. The
non-detections of \msp\ in the 2015 April 12 and 2016 July
6 3.1\,GHz data are, instead, not related to an
unfavourable orbital phase, since in the 2015 data \msp\ has been
detected at 0.7\,GHz and in the 2016 one the observation has been
carried out around inferior conjunction. In the case of the 2015 data,
this is most likely the effect of interstellar scintillation (\S
\ref{subsec:scint}), a phenomenon that can occur at such a high
frequency (see, e.g., Lewandowski et al.\ 2011, who studied the
scintillation parameters of the pulsar PSR\,B0329+54 at 4.8\,GHz). In
the case of the 2016 data, scintillation can be a possible explanation
too. Another possible explanation invokes a time-variable distribution
of the intra-binary plasma, whose effects on the pulsar signal
consequently change with time. If this were the case,
the degree of variability of the intra-binary gas structure would have to be
extremely high, requiring it to change from a situation where the signal
is unaffected for about half orbit at 1.4\,Ghz, to another one where
it embeds the whole binary system with a density high enough to
completely absorb the pulsar signal at 3.1\,GHz. Such dramatic changes
have indeed been observed in the binary RB PSR\,J1740-5340A (D'Amico
et al.\ 2001), where delays of the signal at 1.4\,GHz are observed at
all orbital phases, and the phases at which they occur change substantially from one orbit to another. The companion in that system has a mass of $M_{\rm C}=0.22-0.32\,M_{\odot}$
(Ferraro et al.\ 2003).  The behaviour of the signal delays in this
system are explained with a high degree of instability of the
intra-binary gas structure, such that it can sometimes embed the whole binary, but at other times leaves more than half of the orbit unocculted. \msp\ has a similar
mass companion in a much tighter orbit. It is therefore possible that when the intra-binary
gas is at its maximum size the entire orbit may lie within the innermost
region where the gas density is high enough to completely absorb the
pulsar signal at 3.1\,GHz. We recall (see Table\,\ref{tab:detections})
that the orbital phase range covered by the 2016 July 6
observation begins at $\phi_{\rm orb}=0.76$, i.e. at inferior
conjunction.

Another explanation for the non-detection on 2016 July 6 might be that of
a turn-off of the radio emission, following a transition from a
rotation-powered to an accretion-powered state, implying that \msp\ is a
transitional RB. Such scenario would require the occurrence of two
transitions between 2016 May 24 and 2016 September 9, two epochs when
\msp\ has been detected as a radio pulsar, and the observation of
evidence for the presence of an accretion disk. No long-term change in
the $\gamma$-ray flux is seen in the 12-year monitoring of the source by
the {\em Fermi}-LAT (Paper\,I), which one would expect to see from a RB
undergoing a transition (e.g. Torres et al. 2017, Papitto \& De Martino
2020). Moreover, the analysis of optical data taken between 2016 April 30
and
September 11 at 37 different epochs by Strader et al. (2019) rules
out this scenario. No evidence of optical emission lines due to the
presence of an accretion disk has been found in any of the mentioned
observation, thus implying that \fgl\ remained in its radio pulsar state
between 2016 April and September.

\subsubsection{Pulse brightness vs frequency}
\label{subsec:scint}

\begin{figure*}
  \centering
  \includegraphics[angle=0,width=\textwidth]{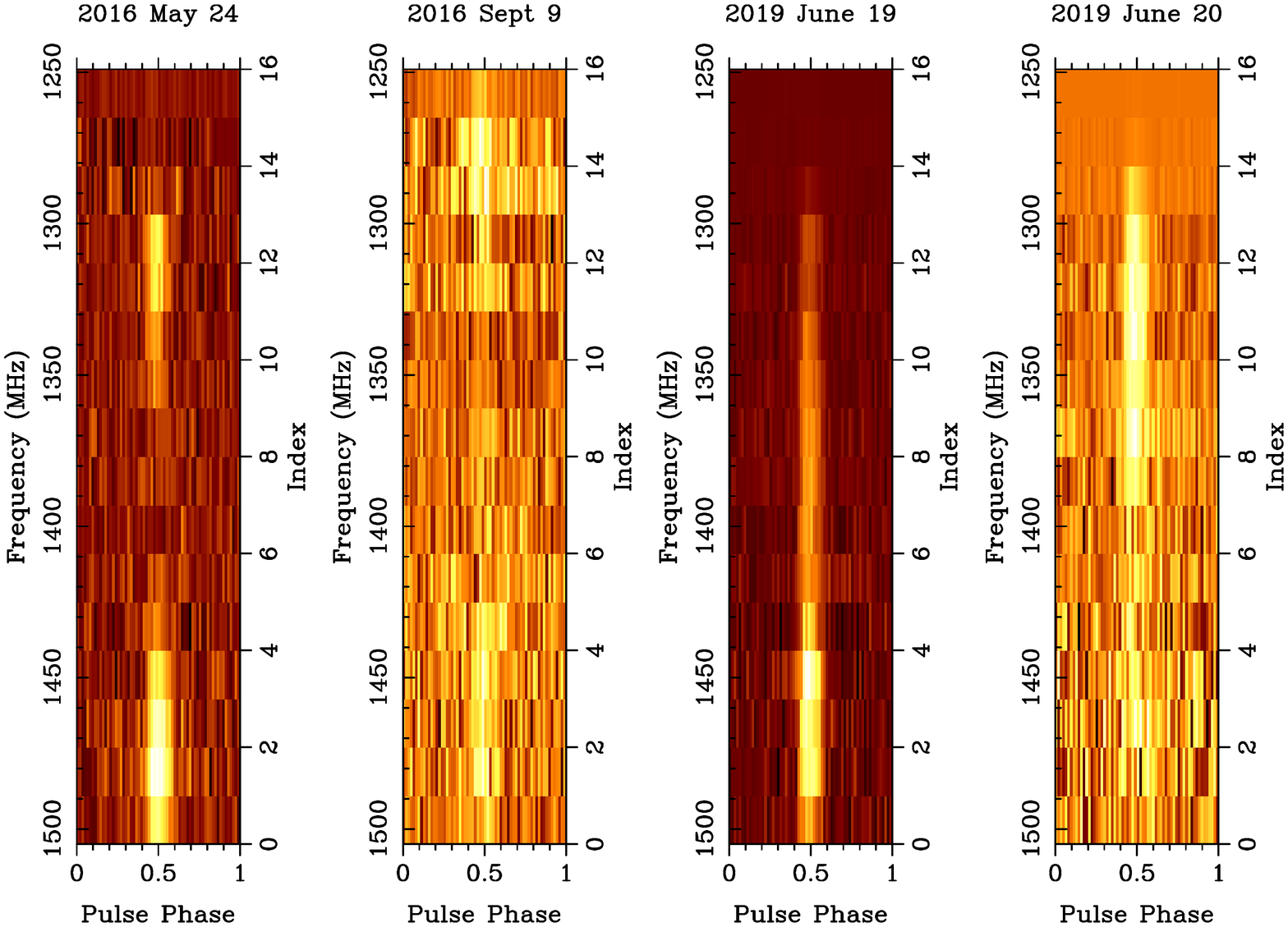}
  \caption{Colour map of the pulse amplitude as a function of the
      pulse phase and frequency channel for all the 1.4\,GHz
      observations where pulses have been detected, left to right:
      2016 May 24, 2016 September 9, 2019 June 19 and 2019 June
      20.  In all panels, the scale on the right-hand side vertical
      axis is the frequency channel. The pulse discontinuity at random
      frequencies is apparent
      across all the observations and is caused by interstellar
      scintillation (see discussion in \S \ref{subsec:scint}). \label{fig:20cm-scint}}
\end{figure*}

Figure\,\ref{fig:20cm-scint} displays a colour map of pulse amplitude
against pulse phase and observing frequency for
1.4\,GHz observations where pulses have been detected. The right panel
of figure\,\ref{fig:Pband-plots} displays the same plot for the
0.7\,GHz data taken on 2015 April 12.  As can be seen, the
pulse brightness is not uniform across the whole band, but peaks in
certain frequency ranges, which vary at random in the four
observations at 1.4\,GHz. Similarly, the pulse brightness also varies
in a random way across the observations. Behaviour of this kind is
typical of interstellar scintillation, which is to be expected at these
frequencies for a dispersion measure of a few tens of pc\,cm$^{-3}$. The observed flux variations are consistent with the values for the
decorrelation bandwidths, $\Delta\nu_{\rm s}=8.8_{-2.9}^{+5.6}$\,MHz
at 1.4\,GHz and $\Delta\nu_{\rm s}=2.1_{-0.7}^{+1.3}$\,MHz at
0.7\,GHz, and the scintillation times, $\Delta t_{\rm
  s}=535_{-53}^{+83}$\,s at 1.4\,GHz and $\Delta t_{\rm
  s}=390_{-39}^{+61}$\,s at 0.7\,GHz, predicted by the NE2001 model
for the Galactic electron distribution along the LOS to \msp{} at the
measured DM.

\begin{figure*}
  \centering
    \includegraphics[angle=0.0,width=\textwidth]{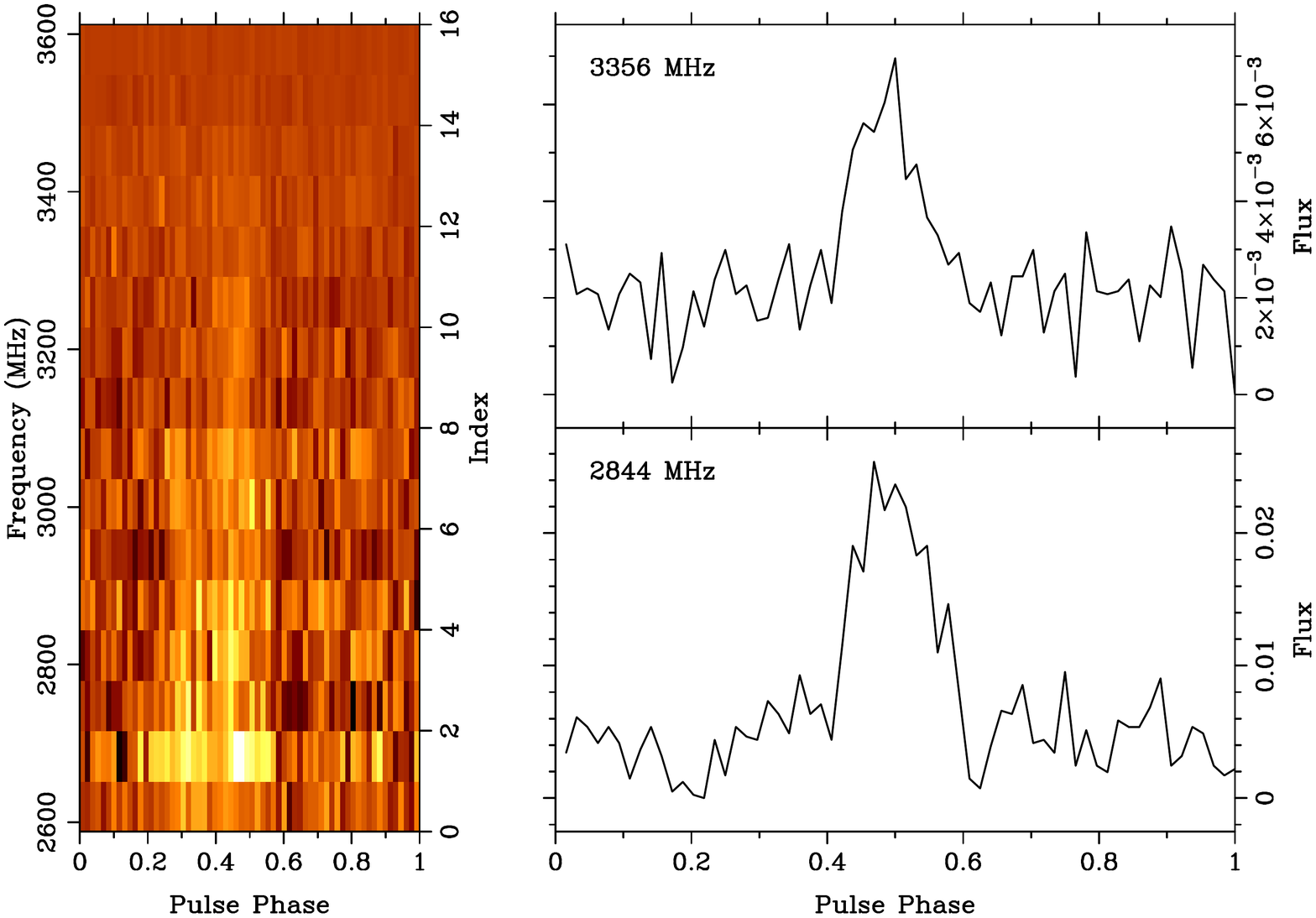}
    \caption{Left: colour map of the pulse amplitude as
      a function of the pulse phase and frequency for the 2016 May
      8 3.1\,GHz observation.  Right: Pulse profile at two
      different frequencies close to the extremes of the 3.1\,GHz where
      pulsations are detectable. \label{fig:sch}}
\end{figure*}

The left-hand panel of Figure\,\ref{fig:sch} displays the same plot
as in Figure\,\ref{fig:20cm-scint} but for the 3.1\,GHz full-orbit observation. As in the
1.4\,GHz data, the pulse brightness is not uniform across the entire
frequency range. The pulse is brighter at lower frequencies
(2600--2800 MHz) although it is still clearly detectable at higher
frequencies (2900--3400 MHz). The NE2001 model predicts a decorrelation bandwidth at this frequency of $\Delta\nu_{\rm
  s}=3.8_{-1.3}^{+2.4}$\,GHz, which implies the visibility of pulses
along a 1\,GHz frequency band at 3.1\,GHz. The two right panels in
Figure\,\ref{fig:sch} display the integrated profiles for each half of
the band, and confirm that the pulsed emission is detectable along the
entire 3.1\,GHz.

\subsection{Pulsar Flux}
\label{subsec:flux}

By using the radiometer equation we obtained a preliminary estimate of
the radio flux density of \msp. To this aim, we considered all
observations where pulses have been detected, thus obtaining the flux
density at three frequencies. For each observation, we considered
those sub-integrations and channels only where pulses were clearly
visible.  We performed this selection by using the interactive routine
{\tt pazi}. We did our selection separately for
sub-integrations and channels, and we adopted an inverse approach: we
removed channels/sub-integrations in which the pulse was bright and
proceeded until the resulting integrated profile showed no evidence of
pulsation. Integration times and bandwidths where the data meet the
above requirement are listed in Table\,\ref{tab:detections}. For those
observations where the signal was detected in separate frequency
sub-bands, the reported bandwidth is the sum of the single signal
bandwidths in each archive. In the considered 3.1\,GHz archive, we did
not discard any sub-integration (\S\,\ref{subsec:eclipses}) nor any
channel (\S\,\ref{subsec:scint}). The resulting archives have been
processed with the routine {\tt pdmp} to obtain the pulses' S/N, whose
values are reported in Table\,\ref{tab:detections}. As discussed in
\S\,\ref{subsec:profs}, the pulse equivalent width, i.e. the width
of an equal height and area rectangle, is $W_{\rm eq}=0.1$
in pulse phase at all considered frequencies.

The radiometer equation also requires the values for the system
temperature $T_{\rm sys}$ and gain $G$ for each receiver. The Parkes
telescope documentation reports a $T_{\rm sys}$ of 40 and 35\,K for the
40cm (0.7\,Ghz) and the 10cm (3.1\,GHz) feed, respectively, of the
10$-$40 receiver, and a $T_{\rm sys}$ of 28\,K for the Multibeam
receiver. The reported gain is $G=1.1$\,Jy/K for all receivers. The
resulting flux densities are reported in
Table\,\ref{tab:detections}. We strongly invite the reader to consider
these values, and their implications, with due care, since is not
known how they are affected by interstellar scintillation (\S\,\ref{subsec:scint}).

\section{Discussion}
\label{sec:discussion}

\subsection{Modelling of radio and $\gamma$-ray light curves}
\label{subsec:align}

In  Paper\,I, the pulsar's orbital parameters and the companion star's
radial velocity amplitude (measured by Strader et al.\, 2019), were
used to constrain the mass of PSR J2039$-$5617, yielding the
constraint that $M_\mathrm{psr} \sin^3 i = 1.04\pm0.05\,M_\odot$ for
an unknown orbital inclination angle $i$.  By fitting models for the
companion star to the optical light curves (LCs), the orbital
inclination was estimated to be $61\degr < i < 78\degr$, which
corresponds to a pulsar mass range of $1.1,M_\odot < M_\mathrm{psr} <
1.6\,M_\odot$.

The detection of radio pulsations from PSR\, J2039$-$5617 (alongside the
$\gamma$-ray pulsations) allows for an additional, independent method
by which to determine the orbital inclination: joint fitting of the
phase-matched radio and $\gamma$-ray LCs.  Under the assumption that
the pulsar was spun-up to its current spin period through the
accretion of stellar material stripped from the companion star, the
spin axis of the pulsar will simultaneously have been aligned with the
orbital axis of the binary system. Hence, the orbital inclination
of the binary system should be equal to the observer angle
$\zeta$ associated with the pulsar (the angle between the observer's
line of sight and the pulsar's spin axis), which can be estimated
through joint fits to the pulsar's radio and $\gamma$-ray LCs.

To align the radio and $\gamma$-ray pulse profiles in phase, we took the 2019 June 19 radio data, which had the highest signal-to-noise ratio, and folded using the $\gamma$-ray ephemeris. As mentioned in Section 3.1, there was still a small linear trend in the pulse phase. We therefore refit the orbital phase to account for this, finding a small offset consistent with the uncertainty from the $\gamma$-ray ephemeris. This offset leads to a negligible phase shift of less than 0.4\% of a rotation, around 12\% of the width of the phase bins adopted for the $\gamma$-ray pulse profile. Along with the best-fitting DM, the epoch, observing location and observing frequency defining the fiducial phase zero (the \texttt{tempo2} parameters TZRMJD, TZRSITE and TZRFRQ, respectively; Hobbs et al. 2006) were used to phase-align the $\gamma$-ray pulse profile to the radio one.

To constrain the value of $\zeta$ for PSR\, J2039$-$5617, two joint fits
to the radio and $\gamma$-ray LCs were conducted.  These fits yielded
not only constraints on $\zeta$, but also on the pulsar's magnetic
inclination angle $\alpha$ (the angle between the pulsar's magnetic
field and spin axes).  The geometric outer gap model (OG; Venter et al.\, 2009),
which is a representation of the physical OG model (Cheng et al.\,
1986a), was used to model the $\gamma$-ray emission for the first of
these fits, while the geometric two-pole caustic model (TPC; Dyks and
Rudak, 2003), which is a representation of the slot gap physical model
(Arons, 1983; Muslimov and Harding, 2003; Muslimov and Harding, 2004),
was used for the second.  For both fits, an empirical single-altitude
hollow cone geometric model (henceforth simply ``Cone''; Story et al.\,
2007) was used for the radio emission.  The observed radio and
$\gamma$-ray LCs were binned with $n_\mathrm{r} = 64$ and $n_\gamma = 30$
equally-spaced phase bins, respectively, and model  LCs binned to match.

We note that the geometric OG and TPC models are based on the retarded vacuum magnetic field structure (Deutsch 1955), while newer kinetic (particle-in-cell) models calculate the magnetic field structure for different assumptions on the plasma density, and therefore cover the entire spectrum from vacuum to force-free (plasma-filled) magnetospheres (e.g., Cerutti et al.\ 2016, Kalapotharakos et al.\ 2018, Philippov \& Spitkovsky 2018).
In the former group of models, $\gamma$-ray emission occurs in regions interior to the light cylinder (where the rotation speed equals that of light in vacuum), while in the latter group, particle acceleration, and therefore $\gamma$-ray emission, is ever more constrained to the equatorial current sheet region close to and beyond the light cylinder as the particle density increases.
While the latter group of models may be more realistic, we opt to use the OG and TPC models in this paper for a number of reasons.
First, the newer models are computationally very expensive, and are typically only solved over a course grid in $\alpha$ and $\zeta$ (and spin-down luminosity).
In contrast, the results presented in this paper for the OG and TPC models benefit from a $1\degr$-resolution in these angles.
Second, these newer models make a number of assumptions, and their physics is still being constrained by data. For example, they invoke scaled-down magnetic fields to make their calculation feasible, as well as a variety of assumptions regarding particle injection rates, and rely on an approximate treatment of the pair production process. Their use may therefore not necessarily lead to statistically improved LC fits (compared to those that result when the OG and TPC models are used), both in the $\gamma$-ray\textendash{}only and joint-fitting contexts.
On the contrary, for some of the newer models, the LC shapes they produce do not seem to be representative of those observed from the $\gamma$-ray pulsar population.
Third, the newer models typically focus on the $\gamma$-ray band and do not take the constraints yielded by the radio models into account.
This means that there is no clear consensus on how the assumptions and simplifications these models rely on affect the predicted radio emission.
In contrast, our joint fits are performed within a single, unified framework, wherein the radio and $\gamma$-ray emission occur within the same magnetic field structure.
This is especially relevant when performing joint LC fits, since it is not only the $\gamma$-ray\textendash{}band goodness of fit that determines which parameter combination is preferred, but also, equally, the radio-band goodness of fit.
Lastly, it is interesting to note that the sky maps (the distribution of radiation vs.\ $\zeta$ and observer phase) are broadly similar for these two groups of models, despite the differences in the mechanisms by which the caustics (which lead to LC peaks as a fixed observer cuts through them) are formed; compare Fig.\ $15 - 17$ of Kalapotharakos et al.\ (2018) and Fig.\ 8 of Philippov \& Spitkovsky (2018) with Fig.\ $16 - 17$ of Venter et al.\ (2009) as well as Fig.\ 10 in this paper. This confirms the foresight by Venter \& Harding (2014), based on prior LC fitting, that newer models should exhibit hybrid behaviour between OG and TPC models.
Comparing the $\gamma$-ray LC for PSR J2039$-$5617 presented here to the published atlas of Kalapotharakos et al.\ (2018), we roughly obtain $\alpha\sim45\degr$ and $\zeta\sim75\degr$ (for a given particle injection rate and spin-down power), which is similar, given all the uncertainties, to the values we find and will discuss later in this section for the OG and TPC models.
Notably, the model atlases associated with the other newer models do not contain LCs with shapes that would fit that observed for PSR~J2039$-$5617, given the different model assumptions that lead to different emissivity distributions.

The parameter space of the OG and TPC models, ${\alpha,\zeta \in
  (0\degr,90\degr)}$, was explored at $1\degr$ resolution, yielding
8,100 candidate pairs of phase-matched radio and $\gamma$-ray model LCs
for each fit.  Since it is unknown \textit{a priori} (at least solely based on
the observed LCs) when (i.e., where in phase) the pulsar's magnetic
field axis is pointing in the observer's direction, an additional
phase shift parameter $\phi_\mu$ was added and explored at a
resolution equal to twice the radio LC bin width of $(1/64)$th of a
rotation.

With this additional parameter implemented, each fit comprised a total
of 259,200 model LC pairs. As a last step, the amplitudes $A_\mathrm{r}$ and $A_\gamma$ of each
of the radio and $\gamma$-ray LCs above the relevant background
levels were
independently adjusted so as to maximize the level of goodness of fit.
In total, then, each pair of model LCs in both fits is associated with
five model parameters: $\alpha$, $\zeta$, $\phi_\mu$, $A_\mathrm{r}$,
and $A_\gamma$. Since the radio and $\gamma$-ray LCs have 64 and 30
bins, the joint fits therefore have $\nu_\mathrm{c} = (64 + 30) - 5 = 89$
degrees of freedom, and the single-band\textendash{}only fits have $\nu_\mathrm{r} = 64
- 4 = 60$ and $\nu_\gamma = 30 - 4 = 26$ degrees of freedom.  

The joint dual-band goodness of fit of each model LC pair $M_\mathrm{c} = (M_\mathrm{r},M_\gamma)$ was
characterized using the scaled-flux standardized (SFS) $\Psi^2_{\Phi,\mathrm{c}}$ goodness-of-fit statistic.  Given an observed pair of
phase-matched radio and $\gamma$-ray LCs, with associated background
levels $b_\mathrm{r}$ and $b_\gamma$, respectively, this statistic is defined as (Seyffert et al. 2016, Seyffert et al. 2020)
\begin{equation}\label{eqn:psqphic}
    \Psi^2_{\Phi,\mathrm{c}}(M_\mathrm{c}) = 1 - \frac{1}{2}\left(
    \frac{\chi^2_\mathrm{r}(M_\mathrm{r})}{\Phi^2_\mathrm{r}} +
    \frac{\chi^2_\gamma(M_\gamma)}{\Phi^2_\gamma} \right),
\end{equation}
where $\chi^2_\mathrm{r}$ and $\chi^2_\gamma$ are the
$\chi^2$ statistics appropriate for the radio and $\gamma$-ray
components of the joint fit, and ${\Phi^2_\mathrm{r} =
  \chi^2_\mathrm{r} (B_\mathrm{r})}$ and ${\Phi^2_\gamma =
  \chi^2_\gamma (B_\gamma)}$ are the squared radio and $\gamma$-ray
scaled fluxes, where ${B_\mathrm{r} = (B_{\mathrm{r},i} =
  b_\mathrm{r})_{n_\mathrm{r}}}$ and ${B_\gamma = (B_{\gamma,i} =
  b_\gamma)_{n_\gamma}}$ are the constant radio and
$\gamma$-ray background-only LCs implied by $b_\mathrm{r}$ and
$b_\gamma$.  The scaled fluxes $\Phi_\mathrm{r}$ and $\Phi_\gamma$
(both squared in Eq.~[\ref{eqn:psqphic}]) measure the total,
pulsar-associated flux contained in the observed radio and
$\gamma$-ray LCs, and are leveraged in
Eq.~(\ref{eqn:psqphic}) to effectively express the component
single-band\textendash{}only deviations ($\chi^2_\mathrm{r}$ and
$\chi^2_\gamma$) in units compatible under addition.

The SFS assigns a goodness-of-fit value of 1 to a `perfect' fit to the data (for which
$\chi^2_\mathrm{r} = \chi^2_\gamma = 0$), and a goodness-of-fit value
of 0 to a fit that is equivalent to assuming the background-only LC pair
$B_\mathrm{c} = (B_\mathrm{r},B_\gamma)$  (for which $\chi^2_\mathrm{r} =
 \Phi^2_\mathrm{r}$ and $\chi^2_\gamma = \Phi^2_\gamma$). A negative value for
$\Psi^2_{\Phi,\mathrm{c}}$ therefore indicates that $M_\mathrm{c}$ is
a worse fit than $B_\mathrm{c}$. The model LC pair $M^\mathrm{sfs}_\mathrm{c}$
for which $\Psi^2_{\Phi,\mathrm{c}} (M^\mathrm{sfs}_\mathrm{c}) = \Psi^2_{\Phi,\mathrm{c,max}}$,
is the model's best fit, and the parameter combination associated with it
constitutes an estimate of the pulsar parameters.

Since $\Psi^2_{\Phi,\mathrm{c,max}}$ is not simply $\chi^2$-distributed, constraints on the parameter estimate due to uncertainties in the LC data are obtained using a Monte Carlo algorithm.
In analogue to the procedure outlined in Avni (1976) for the $\chi^2$ statistic, the goal of this Monte Carlo algorithm is to numerically characterize, via a series of perturbations of the observed LC within its stated flux errors, the distribution of $\Delta\Psi^2_{\Phi,\mathrm{c}} = \Psi^2_{\Phi,\mathrm{c}} (M^\mathrm{sfs}_{\mathrm{c},k}) - \Psi^2_{\Phi,\mathrm{c}} (M^\mathrm{sfs}_\mathrm{c}) \geq 0$, where $M^\mathrm{sfs}_{\mathrm{c},k}$ is the model's best fit in the $k$th iteration (which may or may not be $M^\mathrm{sfs}_\mathrm{c}$), and both $\Psi^2_{\Phi,\mathrm{c}}$ values are calculated with respect to the $k$th-iteration (i.e., perturbed) observed LCs.
Specifically, the goal of the algorithm is to find an estimate $\delta_{3\sigma}$ for this distribution's $3\sigma$ confidence limit.
In the interest of reliability, the algorithm terminates iteration based on a convergence criterion for $\delta_{3\sigma}$ (which is calculated at the end of each iteration): At $200$ iteration intervals (starting from the $400$th iteration), the standard deviation in $\delta_{3\sigma}$ across the preceding $200$ iterations is calculated; if this deviation is less than $5\%$ of $\delta_{3\sigma}$ itself for two consecutive such intervals, iteration is halted.
The final value for $\delta_{3\sigma}$ is then converted into an acceptance contour in parameter space by identifying all the LC pairs for which $\Psi^2_{\Phi,\mathrm{c}} \geq \Psi^2_{\Phi,\mathrm{c,max}} - \delta_{3\sigma}$.
The extent of this contour for each parameter then translates into the desired constraint on that parameter.

The SFS statistic, as compared to the corresponding Pearson's $\chi^2$
statistic ${\chi^2_\mathrm{c}(M_\mathrm{c}) =
  \chi^2_\mathrm{r}(M_\mathrm{r}) + \chi^2_\gamma(M_\gamma)}$, is
better suited to joint fits where the single-band\textendash{}only
best fits associated with the component single-band models correspond
to contradictory estimates for the shared model parameters (in this
case $\alpha$, $\zeta$, and $\phi_\mu$), i.e., where single-band LC fitting yields best-fit parameter estimates that are non-colocated in parameter space.
Joint fits where such non-colocation is present are susceptible to single-band (typically radio) dominance, particularly in cases where the relative errors in one band are much smaller than those in the other.
For example, if the single-band parameter estimates are non-colocated and the relative errors are much smaller for the radio LC than for the $\gamma$-ray LC, the joint fit will be radio dominated, with the best-fit LC pair typically comprised of a good fit to the radio data and a very bad fit to the $\gamma$-ray data.

Seyffert et al.\ (2016) demonstrate that the SFS statistic effectively eliminates single-band dominance in joint fits, and that the best-fit parameter estimates obtained using the SFS statistic converge to those obtained using $\chi^2_\mathrm{c}$ as the respective single-band parameter estimates become more colocated and the error disparity dissipates.
In essence, the SFS statistic yields a compromise solution that typically reproduces the broad LC structure in both bands, despite any error disparity that might exist.

\begin{table}
  \centering
  \caption{Single-band\textendash{}only goodness-of-fit values and parameter estimates for PSR
    J2039$-$5617. Higher values for
    $\chi^2_\nu$ indicate decreased goodness of fit.}
  \label{tab:sb-gof-vals}
  \begin{tabular}{lllllr}
    \hline
    Fit &
    \multicolumn{1}{c}{$\alpha$} &
    \multicolumn{1}{c}{$\zeta$} &
    \multicolumn{1}{c}{$\phi_\mu$} &
    \multicolumn{1}{c}{$\beta$} &
    \multicolumn{1}{c}{$\chi^2_\nu$} \\
    & \multicolumn{1}{c}{($\degr$)} & \multicolumn{1}{c}{($\degr$)} & & \multicolumn{1}{c}{($\degr$)} & \\
    \hline
    Cone & $58^{+3}_{-2}$ & $20^{+3}_{-1}$ & $-0.13$ & $-38^{+1}_{-1}$ & $1.68$ \\
    OG & $71^{+2}_{-2}$ & $29^{+3}_{-2}$ & $-0.09$ & $-42^{+5}_{-4}$ & $6.10$ \\
    TPC & $53^{+10}_{-5}$ & $54^{+4}_{-17}$ & $-0.09$ & \phantom{$-1$}$1^{+9}_{-25}$ & $5.74$ \\
    \hline
    \end{tabular}
\end{table}

\begin{figure*}
    \centering
    \includegraphics[width=140mm]{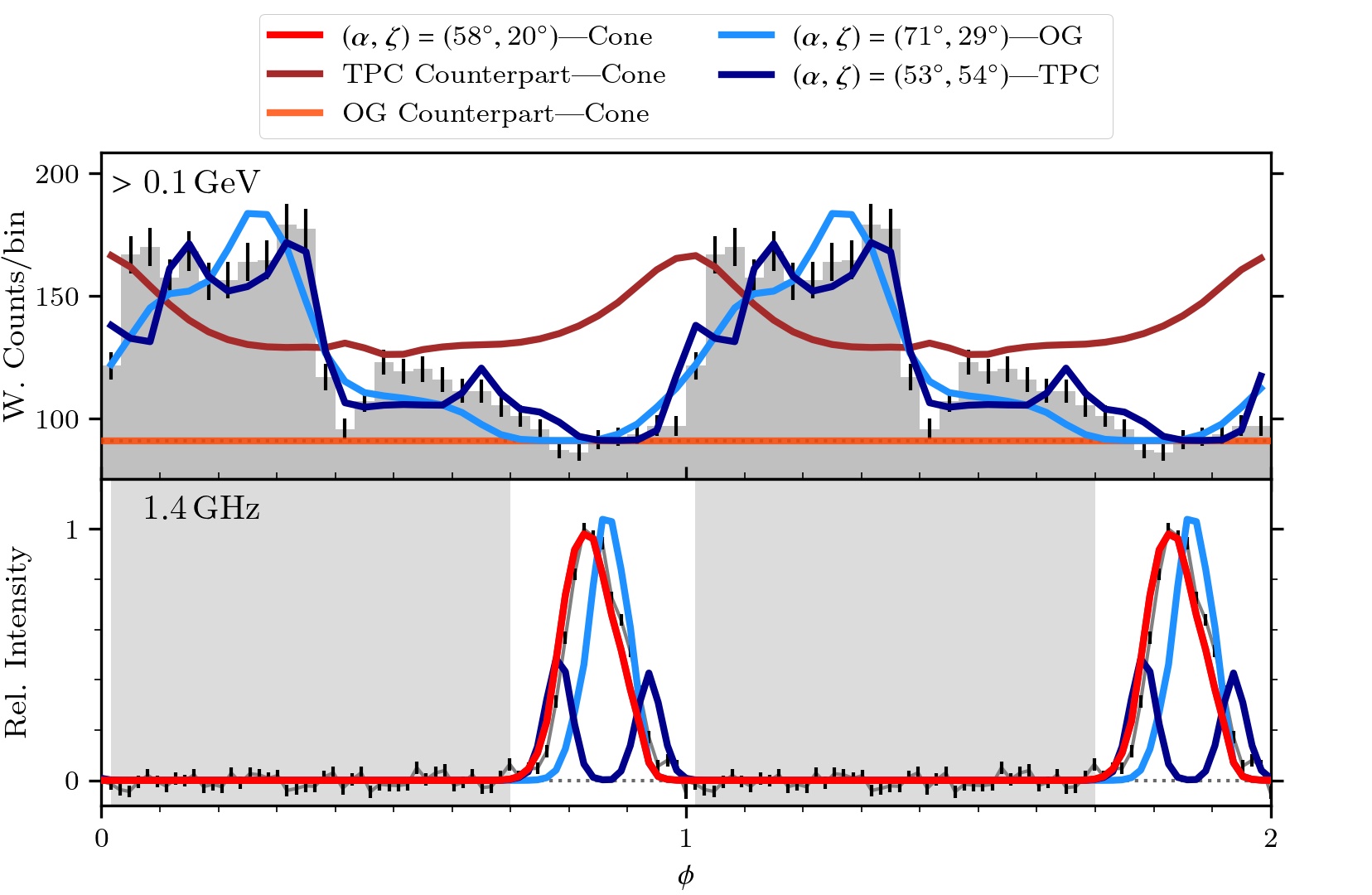}
    \caption{The radio- and $\gamma$-ray\textendash{}only best-fit LCs for the Cone (red; bottom panel), OG (light blue; top panel), and TPC (dark blue; top panel) models, along with their respective $\gamma$-ray and radio counterparts. The observed radio (bottom panel) and $\gamma$-ray (top panel) LCs (both gray with black error bars) are those observed at 1.4\,GHz and at $>0.1$\,GeV, respectively. The uniform radio LC error is taken to be the standard deviation of the intensities in the off-peak region (the gray bands).}
    \label{fig:sb-bflcs}
\end{figure*}

Table~\ref{tab:sb-gof-vals} lists the single-band\textendash{}only parameter estimates, obtained by minimizing $\chi^2$ (or, equivalently $\chi^2_\nu = \chi^2/\nu$; the reduced $\chi^2$ statistic), which are indeed non-colocated in the models' shared parameter space (when comparing the radio-only estimate to each of the $\gamma$-ray\textendash{}only estimates).
The radio- and $\gamma$-ray\textendash{}only best-fit LCs, along with their respective $\gamma$-ray and radio counterparts, are plotted in Figure~\ref{fig:sb-bflcs}.
The radio-only best-fit LC is accompanied by a background-only OG model LC, for which $\chi^2_{\nu,\gamma} = \chi^2/\nu_\gamma = \Phi^2_\gamma/\nu_\gamma = 45.00$, and a TPC model LC that is a worse fit than a background-only LC ($\chi^2_{\nu,\gamma} = 80.40 > \Phi^2_\gamma/\nu_\gamma$).
The $\gamma$-ray\textendash{}only best-fit LC for the OG model is accompanied by a comparatively good radio LC ($\chi^2_{\nu,\mathrm{r}} = \chi^2/\nu_\mathrm{r} = 54.92 \simeq 0.37 \times \Phi^2_\mathrm{r}/\nu_\mathrm{r}$), which resembles the shape of the observed radio LC despite the radio peak occurring too late in phase.
The best-fit LC for the TPC model is accompanied by a somewhat worse radio fit ($\chi^2_{\nu,\mathrm{r}} = 113.28 \simeq 0.75 \times \Phi^2_\mathrm{r}/\nu_\mathrm{r}$), with two radio peaks instead of one.
The comparatively better performance of the OG best fit's counterpart is consistent with the greater agreement between the estimates it and the Cone best fit yield for $\beta = \zeta - \alpha$, the parameter that typically governs peak multiplicity in the Cone model.

\subsubsection{The best joint fits}
\begin{figure*}
    \centering
    \includegraphics[width=140mm]{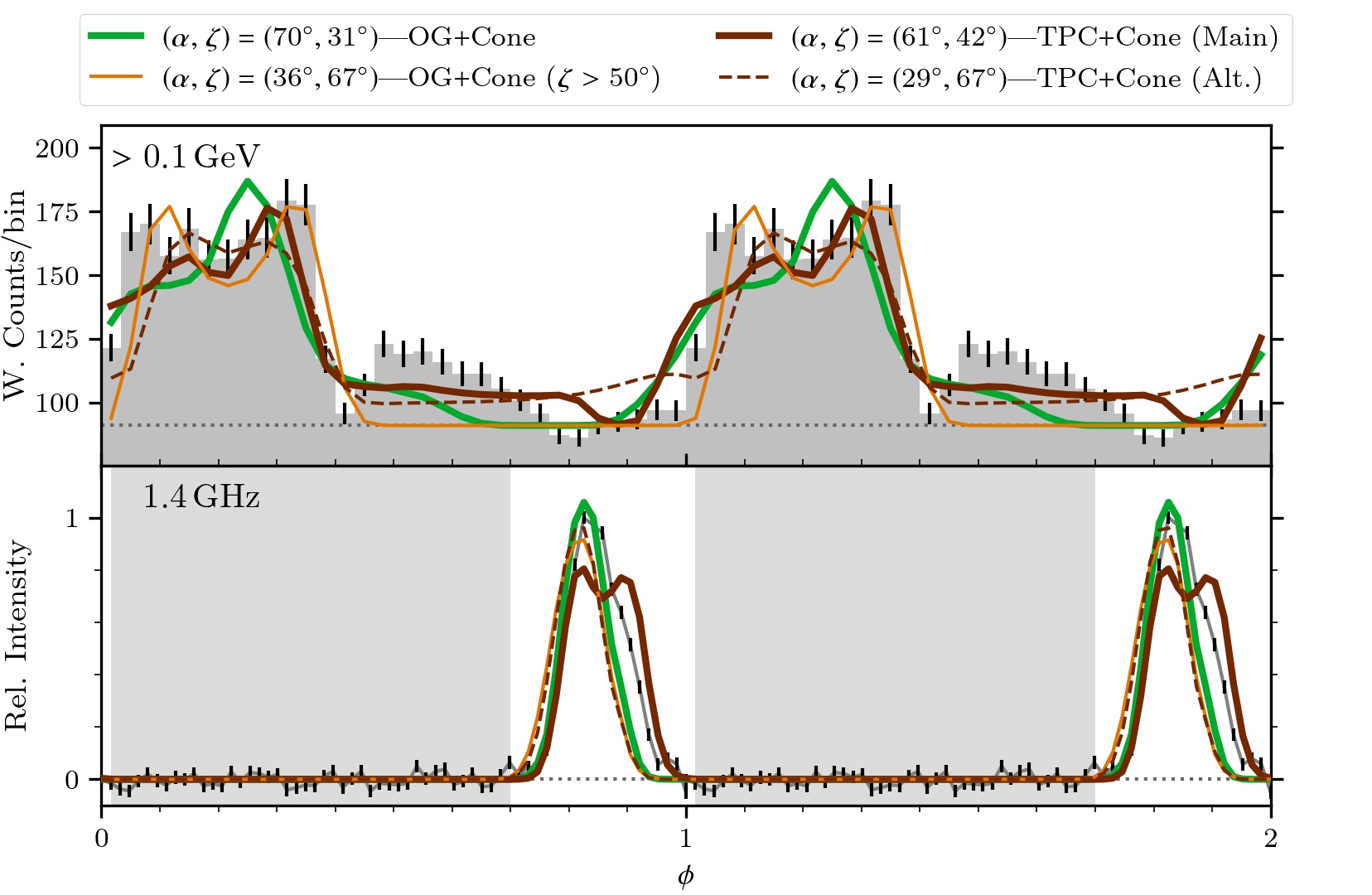}
    \caption{The best-fit LC pairs associated with the OG$+$Cone and
      TPC$+$Cone dual-band models (solid green and solid brown LCs,
      respectively). For the TPC$+$Cone model, the best fit included
      in the alternative contour identified in
      Figure~\ref{fig:sfnmaps}b is also plotted (dashed brown LC). The observed LCs are the same as those in Figure~\ref{fig:sb-bflcs}.
    }
    \label{fig:db-bflcs}
\end{figure*}

\begin{figure*}
    \centering
    \includegraphics[width=\textwidth]{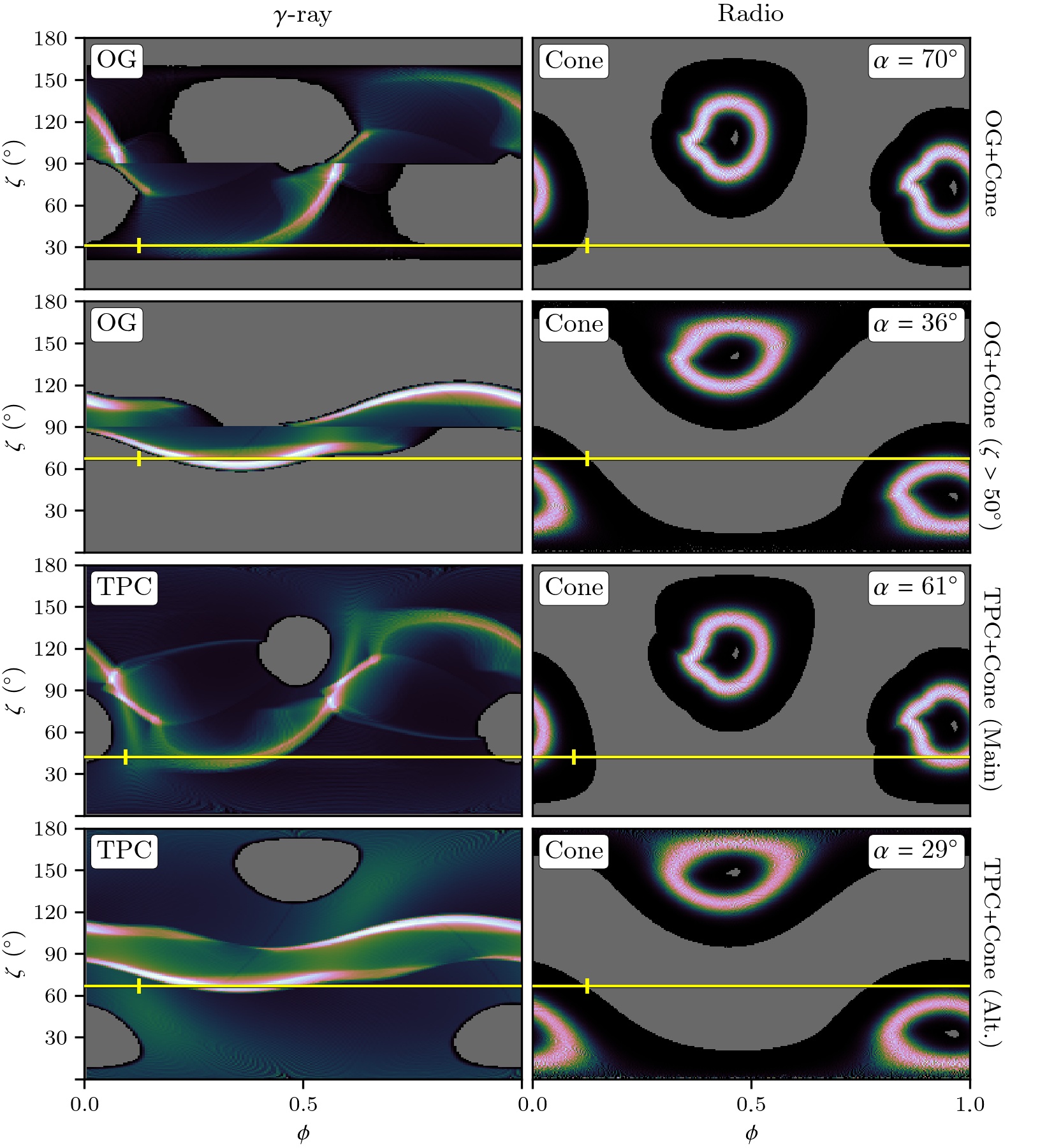}
    \caption{Model radio and $\gamma$-ray emission maps (relative
      intensity as function of rotational phase $\phi$ and observer
      angle $\zeta$) for the four joint fits reported in
      Table~\ref{tab:db-gof-vals}. The grey regions indicate where the
      models predict no pulsar-associated emission, i.e., where only
      background emission will be observed. Each pair (row) of
      emission maps is associated with a single pulsar inclination
      angle $\alpha$, and the LC pairs plotted in
      Figure~\ref{fig:db-bflcs} are each associated with a
      constant-$\zeta$ cut through the appropriate pair of emission
      maps (indicated by the horizontal yellow lines). For each cut,
      the point in model phase that corresponds to 0 in observational
      phase is indicated by a small vertical tick mark (also yellow).}
    \label{fig:phaseplots}
\end{figure*}

\begin{figure*}
    \centering
    \includegraphics[width=\textwidth]{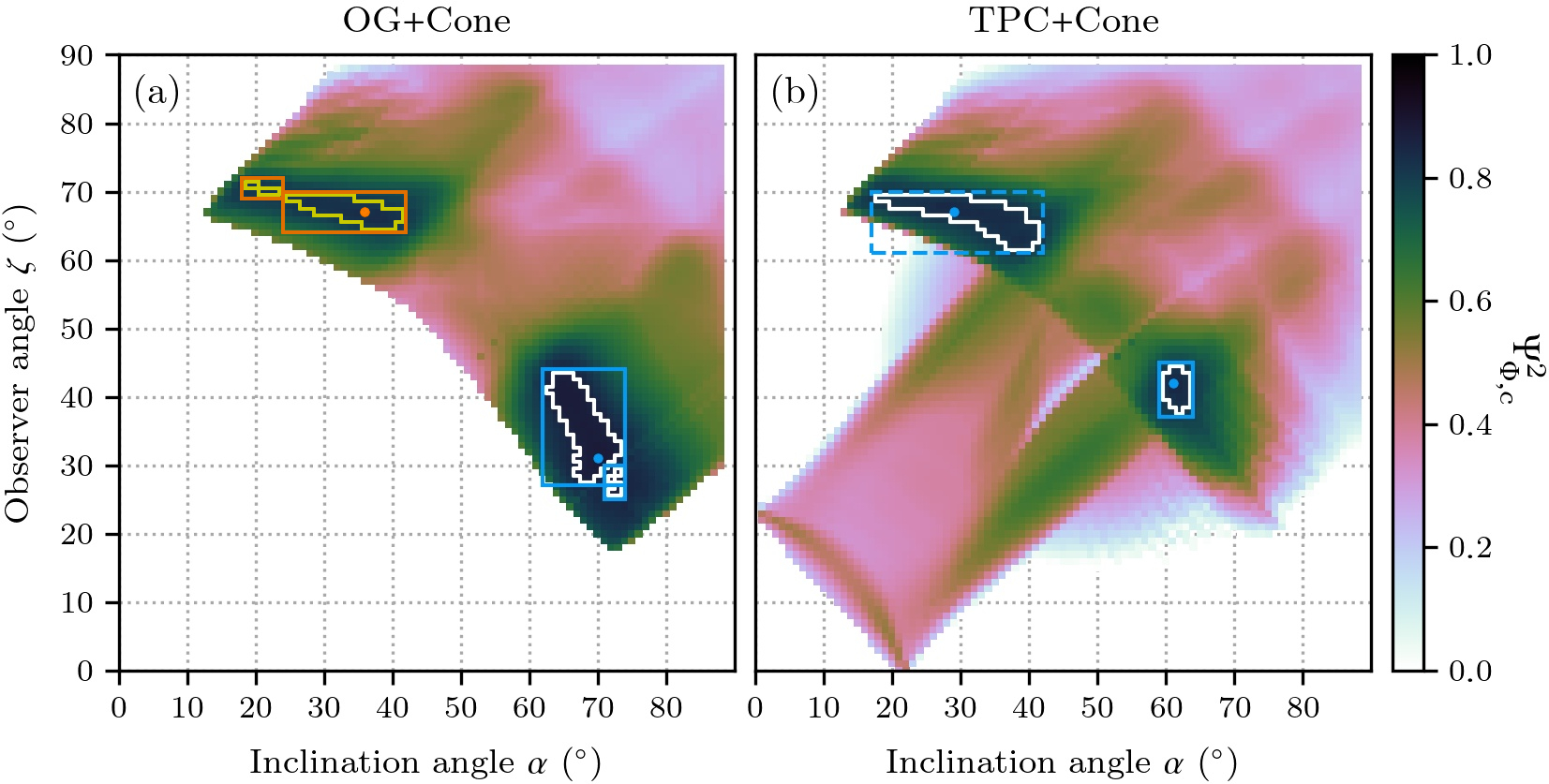}
    \caption{Maps of joint goodness of fit as function of $\alpha$ and
      $\zeta$ (marginalized over $\phi_\mu$) for the OG$+$Cone and
      TPC$+$Cone dual-band models (panels \textit{(a)} and
      \textit{(b)}, respectively) as characterized using
      $\Psi^2_{\Phi,\mathrm{c}}$. Only model LC pairs for which
      $\Psi^2_{\Phi,\mathrm{c}} > 0$ and where both wavebands predict
      pulsar-associated flux, are plotted. In both panels, the white
      contour demarcates the Monte Carlo\textendash{}derived $3\sigma$
      confidence region(s), and each region's best fit is indicated
      with a blue dot. The smaller region for the OG$+$Cone model is
      considered an extension of the larger region, while the
      TPC$+$Cone regions are treated as separate regions. The
      parameter constraints listed in Table~\ref{tab:db-gof-vals} are
      derived from the blue bounding boxes that enclose the various
      regions.}
    \label{fig:sfnmaps}
\end{figure*}

The best-fit LC pair for the OG$+$Cone fit, plotted in green in
Figure~\ref{fig:db-bflcs}, has a joint goodness-of-fit value of
${\Psi^2_{\Phi,\mathrm{c}} = 0.883}$ and a reduced $\chi^2$ value of
${[\chi^2_\mathrm{c}]_\nu = \chi^2_\mathrm{c}/\nu_\mathrm{c} = 6.02}$, as listed
in Table~\ref{tab:db-gof-vals}.  The emission maps associated with
this LC pair are shown in Figure~\ref{fig:phaseplots} (top row).  The
radio-only goodness-of-fit value for this fit is
${\Psi^2_{\Phi,\mathrm{r}} = 1 - \chi^2_\mathrm{r}/\Phi^2_\mathrm{r} =
  0.967}$ (${\chi^2_{\nu,\mathrm{r}} = 5.01}$),
and the $\gamma$-ray\textendash{}only goodness-of-fit value is
${\Psi^2_{\Phi,\gamma} = 1 - \chi^2_\gamma/\Phi^2_\gamma = 0.799}$
(${\chi^2_{\nu,\gamma} = 9.02}$). Notice that, by construction,
$\Psi^2_{\Phi,\mathrm{c}} = (\Psi^2_{\Phi,\mathrm{r}} +
\Psi^2_{\Phi,\gamma})/2$.  These LCs, coupled with the 3$\sigma$
confidence regions shown in white in Figure~\ref{fig:sfnmaps}a,
correspond to a pulsar parameter estimate of ${(\alpha,\zeta) =
((70^{+3}_{-7})\degr,(31^{+12}_{-5})\degr)}$.

Comparing this LC pair to those that correspond to the relevant single-band\textendash{}only fits puts this joint fit, and the compromise it represents, into its proper context: at the cost of some goodness of fit in the radio band (as compared to the radio-only fit; $\chi^2_{\nu,\mathrm{r}}$ increases from $1.68$ to $5.01$), the degree to which the $\gamma$-ray LC is reproduced is increased substantially; from a background-only LC to an LC that is very similar to the $\gamma$-ray\textendash{}only best-fit LC ($\chi^2_{\nu,\gamma}$ decreases from $45.00$ to $9.02$). Or, equivalently (as compared to the $\gamma$-ray only fit), the radio peak's phase is recovered in the joint fit ($\chi^2_{\nu,\mathrm{r}}$ decreases from $54.92$ to $5.01$) at the cost of some goodness of fit in the $\gamma$-ray band ($\chi^2_{\nu,\gamma}$ increases from $6.10$ to $9.02$).

\begin{table*}
  \centering
  \caption{Joint dual-band goodness-of-fit values and parameter estimates for PSR
    J2039$-$5617. Higher values for $\Psi^2_{\Phi,\mathrm{c}}$,
    $\Psi^2_{\Phi,\mathrm{r}}$, and $\Psi^2_{\Phi,\gamma}$ indicate
    increased goodness of fit, while higher values for
    $[\chi^2_\mathrm{c}]_\nu$, $\chi^2_{\nu,\mathrm{r}}$, and
    $\chi^2_{\nu,\gamma}$ indicate decreased goodness of fit.}
  \label{tab:db-gof-vals}
  \begin{tabular}{lllllrrrrrr}
    \hline
    Fit &
    \multicolumn{1}{c}{$\alpha$} &
    \multicolumn{1}{c}{$\zeta$} &
    \multicolumn{1}{c}{$\phi_\mu$} &
    \multicolumn{1}{c}{$\beta$} &
    \multicolumn{1}{c}{$\Psi^2_{\Phi,\mathrm{c}}$} &
    \multicolumn{1}{c}{$[\chi^2_\mathrm{c}]_\nu$} &
    \multicolumn{1}{c}{$\Psi^2_{\Phi,\mathrm{r}}$} &
    \multicolumn{1}{c}{${\chi^2_{\nu,\mathrm{r}}}$} &
    \multicolumn{1}{c}{$\Psi^2_{\Phi,\gamma}$} &
    \multicolumn{1}{c}{${\chi^2_{\nu,\gamma}}$} \\
    & \multicolumn{1}{c}{($\degr$)} & \multicolumn{1}{c}{($\degr$)} & & \multicolumn{1}{c}{($\degr$)} & & & & & & \\
    \hline
    OG+Cone & $70^{+3}_{-7}$ & $31^{+12}_{-5}$ & $-0.13$ & $-39^{+18}_{-8}$ & $0.883$ & $6.02$ & $0.967$ & $5.01$ & $0.799$ & $9.02$ \\
    OG+Cone ($\zeta > 50\degr$) & $36^{+5}_{-16}$ & $67^{+4}_{-2}$ & $-0.13$ & \phantom{$-$}$31^{+21}_{-6}$ & $0.834$ & $11.12$ & $0.924$ & $11.50$ & $0.743$ & $11.52$ \\
    TPC+Cone (Main) & $61^{+2}_{-1}$ & $42^{+2}_{-4}$ & $-0.09$ & $-19^{+2}_{-5}$ & $0.846$ & $16.50$ & $0.859$ & $21.21$ & $0.833$ & $7.53$ \\
    TPC+Cone (Alt.) & $29^{+12}_{-11}$ & $67^{+3}_{-6}$ & $-0.13$ & \phantom{$-$}$38^{+13}_{-16}$ & $0.841$ & $10.46$ & $0.929$ & $10.69$ & $0.753$ & $11.12$ \\
    \hline
    \end{tabular}
\end{table*}

The TPC$+$Cone fit yielded a pair of confidence regions in
$(\alpha,\zeta)$-space: one for which ${\beta <
  0\degr}$ (the main fit) and one for which ${\beta > 0\degr}$.  The
solid brown LCs in Figure~\ref{fig:db-bflcs} are the best fit included in
the former of these regions (enclosed in a solid blue bounding box in
Figure~\ref{fig:sfnmaps}b), and correspond to a parameter estimate of
${(\alpha,\zeta) = ((61^{+2}_{-1})\degr,(42^{+2}_{-4})\degr)}$.
Similarly, the dashed brown LCs in Figure~\ref{fig:db-bflcs} are the best
fit included in the latter of these regions (enclosed in a dashed blue
bounding box in Figure~\ref{fig:sfnmaps}b), and correspond to a
parameter estimate of ${(\alpha,\zeta) =
  ((29^{+12}_{-11})\degr,(67^{+3}_{-6})\degr)}$.  The emission maps
associated with these two fits are shown in the bottom two rows of
Figure~\ref{fig:phaseplots}, and the associated goodness-of-fit values
are listed in Table~\ref{tab:db-gof-vals}.  The overall best fit for
this pair of models (TPC$+$Cone) lies inside the $\beta < 0\degr$
contour.

Comparing the OG$+$Cone and TPC$+$Cone best fits again demonstrates the effect single-band best-fit non-colocation has on the goodness of fit for the joint fits. For the single-band only fits, the TPC model outperforms the OG model ($\chi^2_\nu = 5.74$ vs. $6.10$), but for the joint fits the OG$+$Cone model outperforms the TPC$+$Cone model ($\Psi^2_{\Phi,\mathrm{c}} = 0.883$ vs. $0.846$). Understood in terms of compromise, the greater degree of non-colocation between the TPC and Cone estimates necessitates a more costly compromise than that made in the OG$+$Cone fit.

Since an orbital inclination angle of $i < 50\degr$ (and hence $\zeta
< 50\degr$) implies an unrealistically high pulsar mass of
$M_\mathrm{psr} > 2.4\,M_\odot$, a second OG+Cone fit was conducted,
wherein only model LC pairs for which $\zeta > 50\degr$ were
considered.  The best-fit LC pair thus yielded, which corresponds to a parameter estimate of ${(\alpha,\zeta) =
  ((36^{+5}_{-16})\degr,(67^{+4}_{-2})\degr)}$, is plotted in orange in
Figure~\ref{fig:db-bflcs}, and the associated confidence region (again
derived using a Monte Carlo algorithm) is shown in yellow in
Figure~\ref{fig:sfnmaps}a (enclosed in an orange bounding box).  The
emission maps associated with this fit are shown in the second row of
Figure~\ref{fig:phaseplots}, and the associated goodness-of-fit values
are listed in Table~\ref{tab:db-gof-vals}.  While this fit is worse
than the overall OG$+$Cone best fit, as indicated by the lower
associated $\Psi^2_{\Phi,\mathrm{c}}$ value, it is still a good fit to
the overall structure of the observed LCs since its goodness-of-fit
value is (substantially) larger than 0.

From these fits, since $\zeta > 50\degr$ from the optical fits, we estimate the observer angle to be $\zeta =
(67^{+4}_{-2})\degr$ (from the second OG$+$Cone fit) or $\zeta = (67^{+3}_{-6})\degr$ (from the alternative TPC$+$Cone contour).
Again assuming that $\zeta \equiv i$, these estimates correspond to
mass ranges of $1.17\,M_{\odot} < M_{\rm psr} < 1.46\,M_{\odot}$ and
$1.19\,M_{\odot} < M_{\rm psr} < 1.63\,M_{\odot}$ for the OG$+$Cone and
TPC$+$Cone
models, respectively. The orbital inclination and pulsar mass
estimates obtained here are qualitatively consistent to those found in
 Paper\,I, since a fairly low pulsar mass is preferred in
both cases. As noted in  Paper\,I, orbital inclination estimates
obtained through optical LC models suffer from potentially large
systematic uncertainties, and hence the truly independent estimates
obtained here are an important validation.

An interesting question is whether the constraints on $\alpha$ and $\zeta$, as found by multi-band light curve modelling, may have implications for heating of the MSP's stellar companion.
The stellar surface of the companion is thought to be heated via the pulsar wind (Stappers et al.\ 2001), emission from the intrabinary shock (e.g., Bogdanov et al.\ 
2011, Schroeder and Halpern 2014) or particles ducted from the shock along magnetic field lines toward the surface (Romani and Sanchez 2016, Sanchez and Romani 2017), or directly via the pulsed $\gamma$-ray
emission from the pulsar. 
Pulsed magnetospheric $\gamma$ rays typically comprise only $\sim10\%$ of the energy budget of MSPs (Abdo et al.\ 2013). In
this case, $\alpha$ and $\zeta$ are important in determining the
fraction of $\gamma$ rays intercepted by the companion. The companion subtends an angle of $\sim2\tan^{-1}(1/5) \approx 22^\circ \lesssim \alpha$. Thus, if the $\gamma$ rays are isotropically radiated within the opening angle $\sim\alpha$ and spin and orbital axes are aligned, a fraction of $\sim 5\%$ of the total
$\gamma$-ray emission will intercepted by the companion at any given orbital phase. Yet, a minor misalignment of spin and orbital axes, by $\sim10^\circ$, would imply that most of the $\gamma$ rays would
miss impacting the companion.  
On the other hand, the pulsar wind may be energetically dominant when considering heating of the companion. The pulsar wind Poynting flux is also anisotropic, and depends on $\alpha$ (e.g., Tchekhovskoy et al.\ 2016). Since the X-ray double-peaked light curve phasing suggests a shock orientation concave toward the MSP (Wadiasingh et al.\ 2017), the energetics is largely determined by the opening angle of the shock, which in turn will depend on $\alpha$ and other factors in pressure balance detailed in Wadiasingh et al.\ (2018). 
The efficiency of heating depends on the opacity in the photosphere, its ionization state and consequently the spectrum of the impinging radiation field. X-rays from the shock could be more efficient in heating the companion if the ionization fraction is low/moderate, while $\gamma$ rays could penetrate deeper into the companion atmosphere and affect it hydrostatically (e.g., London et al.\ 1981, Ruderman et al.\ 1989). Modelling such irradiated atmospheres is a highly nontrivial problem.

\subsection{Pulsar distance}
\label{subsec:distance}

Since its $\gamma$-ray discovery, the \fgl\ distance has been
unknown. Based upon the limits on the hydrogen column density $N_{\rm
  H}$ derived from the fits to the {\em XMM-Newton} spectrum, Salvetti
et al.\ (2015) estimated a distance $d\la0.9$\,kpc, consistent with the
limits they derived from a colour-magnitude analysis of the putative
MSP companion star ($0.2\la d \la 0.9$\,kpc).  From the stellar proper
motion reported in the NOMAD catalogue (Zacharias et al.\ 2005), and
assuming the median of the MSP transverse velocity distribution,
Salvetti et al.\ (2015) inferred a new tentative distance range, which
becomes $0.15\la d \la 2.77$\,kpc after accounting for the MSP
velocity standard deviation.  Repeating the same exercise with the
more recent stellar proper motion from the {\em Gaia} DR2 catalogue,
$\mu_{\alpha}\cos\delta=4.21\pm0.29$ mas yr$^{-1}$ and
$\mu_{\delta}=-14.93\pm0.26$ mas yr$^{-1}$ ({\em Gaia} Collaboration
2018), implies a slightly narrower distance range $0.28\la d \la 2.54$
kpc.  More recently, by modelling the optical light curve of the
companion star, but with no {\em a priori} knowledge of the system's
mass ratio $q$, Strader et al.\ (2019) derived a larger distance of
$3.4\pm0.4$\,kpc.  In  Paper\,I, the modelling of the optical light
curve based on newly acquired data and the knowledge of the system
mass ratio from the $\gamma$-ray and optical mass functions gives
1.7$\pm$0.1 kpc.

With the detection of radio pulsations, the DM provides another way to
estimate the source distance, which, however, depends on the assumed
electron density model. The DM of $24.57\pm0.03$\,pc\,cm$^{-3}$ implies a
distance $d=1.708\pm0.004$\,kpc using the YMW16 model\footnote{{\tt
http://www.atnf.csiro.au/research/pulsar/ymw16/}}, where the error is
only statistical, which becomes as low as $d=0.9\pm 0.2$\,kpc using the
NE2001 model\footnote{{\tt https://www.nrl.navy.mil/rsd/RORF/ne2001/}}.
The former value is more consistent with the distance of
$2.5^{+3.3}_{-0.9}$\,kpc inferred from a preliminary optical parallax
measurement ($\sim 2\sigma$) of the \msp\ companion star given in the
{\em Gaia} DR2 catalogue,
whereas the latter is closer to the estimates based on the X-ray
observations and on the colour-magnitude analysis. The factor of two
discrepancy between the two DM-based values can be smoothed by a
more realistic uncertainty estimate for the distances computed through
the YMW16 model, for which only the statistical error on the DM is
accounted for, whereas for those computed through the NE2001 model, a
nominal 20\% systematic uncertainty is taken into account, although a
factor of 2 higher uncertainty is conservatively assumed in some cases
(e.g., Abdo et al.\ 2013).  To quantify the unknown systematic
distance uncertainty from the YMW16 model, we have compared the
DM-based distances\footnote{{\tt
    https://www.atnf.csiro.au/research/pulsar/psrcat/}} ($d_{\rm
  YMW16}$) with the parallactic distances\footnote{{\tt
    http://hosting.astro.cornell.edu/research/parallax/}} ($d_{\pi}$),
assumed as a reference, on a sample of 134 pulsars and computed the
fractional difference between the two quantities $(d_{\rm
  YMW16}-d_{\pi})/d_{\pi}$. To filter out obvious outliers we rejected
entries where the absolute value of the fractional difference was
larger than 1. We found that the mean of the distribution is 0.12,
with a standard deviation of 0.40.  Our independent uncertainty
estimate is in very good agreement with that obtained by Yao et
al.\ (2017) from an analogous comparison based on a different sample
of 189 pulsars with either parallactic or other non DM-based distance
measurements. Therefore, we assume a realistic uncertainty of 40\% on
the YWM16 DM-based distance.  Under this assumption, the
\msp\ distance would then be $d=1.7\pm0.7$ kpc, which would be
marginally consistent with that obtained from the NE2001 model. The value
obtained in  Paper\,I is larger than what we derived from the DM
and the NE2001 model and more consistent with that derived from the
YWM16 model.  An improved, model-free, measurement of the optical
parallax of the \msp\ companion star in one of the next {\em Gaia}
releases, with DR3 expected in Fall 2021, would hopefully provide an
independent confirmation of the presumedly most likely distance value.

\subsection{Origin of the radio eclipses}

As discussed in \S\,\ref{subsec:eclipses}, the most likely
explanation for the radio eclipses at 1.4\,GHz is attributed to the
presence of intra-binary gas.  In the case of hot intra-binary gas,
the high temperature would imply a very high degree of ionisation,
hence a large density of free electrons that rapidly enhance the
frequency-dependent signal dispersion in the medium close to the
pulsar surface.  This effect is not corrected by the signal
de-dispersion (\S \ref{subsec:detection}) though, which is applied at
a fixed DM value that only accounts for the free electron density in
the interstellar medium (ISM). This hot gas would most likely originate from the diffusion
of the external layers of the companion star ablated by the pulsar
wind, a phenomenon which occurs in RB systems (Rasio et al.\ 1991).
However, the lack of appreciable delays in the radio signal
propagation at 1.4 GHz at the edges of the eclipse (Figure 2) is
also compatible with an alternative hypothesis of a relatively low
ionisation degree of the intra-binary gas. A sanity check for the cold
gas scenario is the investigation of possible variations of the
dispersion measure along the orbit. 

\begin{figure}
  \centering
  \includegraphics[angle=0,width=\columnwidth]{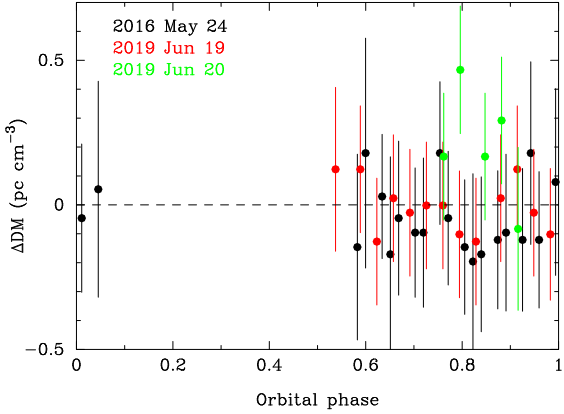}
  \caption{Dispersion measure variations $\Delta$DM versus orbital phase
    $\phi_{\rm orb}$ for the 2016 May 24 (black), 2019 June 19
    (red) and 20 (green) observations. Each data point
    corresponds to a phase bin of 0.024 orbits (8\,minutes), and is the
    difference between the DM in the phase bin and the DM value of the
    observation as reported in Table\,\ref{tab:detections}. Vertical bars
    represent twice the uncertainty on $\Delta$DM. Points have only been
    included if we consider the pulsed to have been detected in a visual
    inspection of the associated integrated profile.
    \label{fig:dm-variations}}
\end{figure}

We considered the three observations
at 1.4\,Ghz where \msp\ has been observed for a large portion of the
orbit, namely the ones taken on 2016 May 24 and the two from
2019. In the corresponding archives, we summed sub-integrations by
groups of six, so that the new ones have sub-integrations of one minute,
then split these into separated sub-archives with eight sub-integrations
each. In this way we obtained archives that span 8\,minutes along the
orbit, i.e. 2.4\% of the orbit. The corresponding orbital phase was
determined from the refined value for the time of ascending node passage
for the observation (\S\,\ref{subsec:detection}), and the DM obtained by
processing these sub-archives with {\tt pdmp}. We only included phase
bins in which we considered the pulse profile to have been detected by
visual inspection of the sub-archive integrated profile.
Figure\,\ref{fig:dm-variations} plots the variation of the DM for the
three mentioned observations with different colors. Each point
represents the difference between the DM value for each orbital phase
bin and the value for the corresponding whole observation reported in
Table\,\ref{tab:detections}. Vertical bars correspond to twice the
uncertainty of the DM difference, i.e. the 90\% confidence level
error. All plotted points except two from the 2019 June 20
observation, where the detected pulses are weakest, are consistent with zero DM variation. The mean
value for the absolute DM variation $\langle|\Delta DM|\rangle$ and the relative
standard deviation $\sigma_{|\Delta DM|}$ are $\langle|\Delta
DM|\rangle=$0.116\,pc\,cm$^{-3}$ and $\sigma_{|\Delta
  DM|}=$0.052\,pc\,cm$^{-3}$ in the 2015 May 24
observation, $\langle|\Delta DM|\rangle=$0.072\,pc\,cm$^{-3}$ and $\sigma_{|\Delta
  DM|}=$0.051\,pc\,cm$^{-3}$ in the 2019 June 19 observation, and
$\langle|\Delta DM|\rangle=$0.24\,pc\,cm$^{-3}$ and $\sigma_{|\Delta
  DM|}=$0.13\,pc\,cm$^{-3}$ in the 2019 June 20 observation. These quantities are consistent with zero at the 2$\sigma$ level.
  
Therefore, we can set a 90\% confidence upper limit of
$\sim$0.4\,pc\,cm$^{-3}$ on the maximum amplitude of any orbital phase
dependent variations in the DM of the pulsar signal. On one side, this
is in agreement with the hypothesis of the presence of a mostly cold
intra-binary gas, but, on the other side, that cannot exclude the
occurrence of a low density fully ionised gas. Observations over a
larger instantaneous bandwidth are needed to improve the sensitivity
to DM variations and possibly constrain the thermodynamical status of
the intra-binary gas. In both cases, according to the as yet
unconstrained size and density distribution of the intra-binary gas,
when the pulsar is close enough to the superior conjunction the radio
signal can travel an optically thick path and it is simply absorbed by
the intervening medium. If its size were up to a few times the size of
the Roche Lobe
of the companion star, then the cloud might be large enough to embed the
whole system, depending on its exact morphology and on the gas
confinement status.  In this case, for a high enough gas density, the
eclipse of the radio signal might be total, which is not the case.

What is in common between the two scenarios is the apparent sudden
transition at the edges of the signal eclipses. The 2019 June 19
observation (Figure\,\ref{fig:fullorbitsL}, mid panel)  shows a
signal behaviour that is fully consistent with what is seen in the
above-mentioned data.  This fact is consistent with the picture that
requires an intra-binary gas whose structure is not perfectly stable in time, but which provides a high
optical depth to the propagating signal.

\section{Summary}
\label{sec:summary}

Through Parkes observations obtained in 2016 we discovered radio
pulsations at 0.7, 1.4 and 3.1\,GHz from the former RB candidate
\fgl\ following its detection as a $\gamma$-ray pulsar (\msp) with
spin period $P_{\rm s}\sim 2.6$ ms ( Paper\,I). At 0.7, 1.4 and 3.1\,GHz
band, the radio signal consists of a broad single peak.
We found clear evidence of eclipses of the radio
signal at 1.4\,GHz for about 50\% of the orbit at pulsar superior
conjunction, which proves that \msp\ features the same radio
phenomenology expected for an RB. The origin of the eclipses cannot be
unambiguously determined from the available data, which
can accommodate absorption of the radio signal by both a cold or a hot
intra-binary gas.  From our radio observations, we provide the first
direct measurement of the dispersion measure
(DM\,=\,$24.57\pm0.03$\,pc\,cm$^{-3}$), corresponding to a distance of
$d=0.9\pm 0.2$\,kpc, assuming the NE2001 model, compatible at
1$\sigma$ with that obtained from the YWM16 model,
$d=1.7\pm0.7$\,pc\,cm$^{-3}$, after a realistic error treatment. 
We also measured the pulsar flux density at all the three observing
frequencies.
A comparison between our 2016 and 2019 1.4\,GHz observations of
\msp\ did not unveil any long-term variation in the pulsar radio
emission, whereas {\em Neil Gehrels Swift Observatory} (hereafter {\em
Swift}) and {\em Fermi} observations showed that
the X and $\gamma$-ray fluxes were stable in the time span from June
2017 to May 2018. We will continue our monitoring observations of
\msp\ in both radio and X-rays to look for possible long-term
state changes in the source flux and determine whether this is one of
the very rare transitional RBs. 
Finally, we matched the (zero) phases of the radio and $\gamma$-ray pulse profiles
finding that the radio pulse leads the
main $\gamma$-ray pulse. We also jointly fitted radio and $\gamma$-ray
pulses against two geometric models, namely the outer gap (OG) and the
two-pole caustic (TPC) ones, and from both models we obtain values for the
magnetic field inclination and observer angles, namely
$(\alpha,\zeta)=({36^\circ}^{+5}_{-16},{67^\circ}^{+4}_{-2})$ and
$(\alpha,\zeta)=({29^\circ}^{+12}_{-11},{67^\circ}^{+3}_{-6})$,
for the OG and TPC, respectively.  Assuming that the pulsar spin axis is aligned to the
orbital axis, i.e.  $\zeta\equiv i$, the light curve modelling
gives an independent measurement of the latter, from which we derive
the ranges for the pulsar mass $M_{\rm psr}$ $1.3\,M_{\odot} < M_{\rm
  psr} < 1.5\,M_{\odot}$ and $1.3\,M_{\odot} < M_{\rm psr} <
1.6\,M_{\odot}$ for the OG and TPC models, respectively. These ranges
are in qualitative agreement to those reported in  Paper\,I.

We conclude by remarking that \msp\ is now one
of a handful of BWs/RBs for which the discovery of an optical/X-ray
periodic flux modulation paved the way to the detection of
$\gamma$-ray/radio pulsations, after the BW PSR\, J1311$-$3430
(Pletsch et al.\ 2012) and the RB PSR\, J2339$-$0533 (Ray et
al.\ 2014). With only a minority of BWs/RBs
detected as X-ray pulsars, the search for X-ray pulsations from
\msp\ is now one of the next steps.  The case of \msp\ confirms the validity of the multi-wavelength
approach in the identification of BW/RB systems and spurs systematic searches (e.g., Braglia et al.\ 2020).  New BW/RB
candidates singled out through the detection of  optical/X-ray
flux modulations will hopefully be confirmed in the next years once
radial velocity measurements provide the values of the orbital
parameters to ease blind radio/$\gamma$-ray periodicity searches.


\section*{Acknowledgements}
We credit the contributions of A. Harding and T. Johnson on the analysis of the pulsar radio and $\gamma$-ray light curves, which paved the way to the modelling reported in this paper. The Parkes radio telescope is part of the Australia Telescope National Facility that is funded by the Australian Government for operation as a National Facility managed by CSIRO. We acknowledge the use of public data from the {\em Swift} data archive. This work is based on the research supported wholly / in part by the National Research Foundation of South Africa (NRF; Grant Numbers 87613, 90822, 92860, 93278, and 99072).  The Grantholder acknowledges that opinions, findings and conclusions or recommendations expressed in any publication generated by the  NRF  supported  research  is  that  of  the  author(s),  and  that  the  NRF  accepts  no  liability  whatsoever  in  this regard.
The \textit{Fermi} LAT Collaboration acknowledges generous ongoing support from a number of agencies and institutes that have supported both the development and the operation of the LAT as well as scientific data analysis. These include the National Aeronautics and Space Administration and the Department of Energy in the United States, the Commissariat \`a l'Energie Atomique and the Centre National de la Recherche Scientifique / Institut National de Physique Nucl\'eaire et de Physique des Particules in France, the Agenzia Spaziale Italiana and the Istituto Nazionale di Fisica Nucleare in Italy, the Ministry of Education, Culture, Sports, Science and Technology (MEXT), High Energy Accelerator Research Organization (KEK) and Japan Aerospace Exploration Agency (JAXA) in Japan, and the K.~A.~Wallenberg Foundation, the Swedish Research Council and the Swedish National Space Board in Sweden. Additional support for science analysis during the operations phase is gratefully acknowledged from the Istituto Nazionale di Astrofisica in Italy and the Centre National d'Etudes Spatiales in France. This work performed in part under DOE Contract DE- AC02-76SF00515.
C.J.C. acknowledges support from the ERC under the European Union's Horizon 2020 research and innovation programme (grant agreement No. 715051; Spiders).
A.R. gratefully acknowledges financial support by the research grant ``iPeska'' (P.I. Andrea Possenti) funded under the INAF national call Prin-SKA/CTA approved with the Presidential Decree 70/2016.

{\it AC, RPM, AP, MB, ADL, AB, AR would like to thank and commemorate
Prof. Nichi D’Amico, who prematurely passed away while this work was
being finalised. Prof. D’Amico has been widely appreciated during his
entire career, in particular as Director of the Sardinia Radio Telescope
project and as President of the Italian National Institute for
Astrophysics, and he represented a leading figure in the radio pulsar
field. He has been a guide for all scientists who had the honour
and privilege of collaborating with him, and his advice and indications
have been very precious and fundamental for the scientific growth of
some authors of this work.}



\section*{Data availability}

The data used in this work are all available in public archives. See references in Section\,\ref{sec:obs}.











\bsp	
\label{lastpage}
\end{document}